\documentclass[useAMS,usenatbib]{mn2e}
\usepackage{graphicx,setspace,xcolor}

\title[The Gemini NICI Planet-Finding Campaign: Asymmetries in the HD
141569 disc]{The Gemini NICI Planet-Finding Campaign: Asymmetries in the HD 141569 disc}
\author[Beth Biller et al.]{Beth
  A. Biller$^{1}$\thanks{E-mail:bb@roe.ac.uk}, Michael C. Liu$^{2}$,
  Ken Rice$^{1}$, Zahed Wahhaj$^{3}$, Eric Nielsen$^{4,5}$, \and
  Thomas Hayward$^{6}$, Marc Kuchner$^{7}$, Laird M. Close$^{8}$, Mark
  Chun$^{9}$, Christ Ftaclas$^{2}$, \and Douglas W. Toomey$^{10}$ \\
$^{1}$Institute for Astronomy, University of Edinburgh, Blackford Hill, Edinburgh EH9 3HJ, UK \\
$^{2}$Institute for Astronomy, University of Hawaii, 2680 Woodlawn Drive, Honolulu, HI 96822, USA \\
$^{3}$European Southern Observatory, Alonso de C\'ordova 3107,
Vitacura, Casilla, 19001, Santiago, Chile \\
$^{4}$Kavli Institute for Particle Astrophysics and Cosmology,
Stanford University, Stanford, CA 94305, USA \\
$^{5}$SETI Institute, Carl Sagan Center, 189 Bernardo Avenue, Mountain
View, CA 94043, USA \\
$^{6}$Gemini Observatory, Southern Operations Center, c/o AURA, Casilla 603, La Serena, Chile \\
$^{7}$NASA Goddard Space Flight Center, Exoplanets and Stellar Astrophysics Laboratory, Greenbelt, MD, 20771, USA \\
$^{8}$Steward Observatory, University of Arizona, 933 North Cherry Avenue, Tucson, AZ, 85719, USA \\
$^{9}$Institute for Astronomy, University of Hawaii, 2680 Woodlawn Drive, Honolulu, HI, 96822, USA \\
$^{10}$Mauna Kea Infrared, LLC, 21 Pookela St., Hilo, HI, 96720, USA \\
}

\begin{document}

\date{}

\pagerange{\pageref{firstpage}--\pageref{lastpage}} \pubyear{2002}

\maketitle

\label{firstpage}

\begin{abstract}
We report here the highest resolution near-IR imaging to date of the HD 141569A disc taken as part of the
NICI (near-IR coronagraphic imager) Science Campaign. 
We recover 4 main features in the NICI images of the HD 141569 disc
discovered in previous HST imaging: 1) an inner ring / spiral feature.
Once deprojected, this feature does not appear circular.
2) an outer ring which 
is considerably brighter on the western side compared to the eastern
side, but looks fairly circular in the deprojected image.
3) an additional arc-like feature between the inner and outer ring
only evident on the east side. In the deprojected image,
this feature appears to complete the circle of the west side inner
ring and 4) an evacuated cavity from 175 AU inwards.  Compared to the
previous HST imaging with relatively large coronagraphic inner working
angles (IWA), the NICI coronagraph allows imaging down to an IWA of 0.3''.
Thus, the inner edge of the inner ring/spiral feature is well resolved
and we do not find any additional disc structures within 175 AU.
We note some additional asymmetries in this system.  Specifically, 
while the outer ring structure looks circular in this deprojection,
the inner bright ring looks rather elliptical.  
This suggests that a
single deprojection angle is not appropriate for this system
and that there may be an offset in inclination between the two ring / spiral
features.  We find an offset of 
4$\pm$2 AU between the inner ring and the star center, potentially
pointing to unseen inner companions.
\end{abstract}


\section{Introduction}

During the process of formation, planets can sculpt complex structures in
their natal discs, observable both at the transitional and debris disc 
stage.  Thus, features such as gaps and spirals within such discs are
often considered signposts of ongoing or recently completed planet
formation.  However, complex structures might form with no
planets present  -- for instance, \citet[]{Lyr13} recently found
that dust-gas interactions in debris discs with remnant gas may also
form sharp features in discs, such as narrow, eccentric rings.
Thus, considerable care in 
data analysis and modeling is necessary to attempt to elucidate the 
origins of imaged structure in any given disc.

One disc which has already proven to be a vital laboratory of in-situ 
planet-formation and disc structure is that around HD 141569A.  
HD 141569A is a B9.5 Ve young pre-main sequence star \citep[]{Mer04}
116$\pm$8 pc from the Sun \citep[]{vLe07}
with co-moving M2 and M4 companions that form a binary pair
$\sim$7.5'' from A \citep[]{Wei00}.  
All three stars show signs of youth and \citet[]{Wei00} find an age for the system of 5$\pm$3 Myr.
The HD 141569A disc is one of the largest
imaged discs known to date, both in projection on the sky, with a
radial extent along its major axis of $>$450 AU (4'' on the sky).
The disc has been variously classified as a debris disc and a transitional
disc, as it still has some remnant gas.  It is now generally 
considered an example of a very young debris disc.  

HST and ground based studies in both optical and near-IR wavelengths
have revealed complex structure in scattered light in this disc
\citep[]{Aug99, Wei99, Mou01, Boc03, Cla03}:
\begin{enumerate}
\item An outer ring or tightly wound spiral at $\sim$400 AU (adopting a
  distance of 116 pc) along the major axis \citep[]{Aug99, Wei99,
    Mou01, Boc03, Cla03},
\item An inner ring or tightly wound spiral at $\sim$245 AU along the major axis
\citep[]{Wei99, Mou01, Cla03}.  While initially interpreted as
rings, the inner and outer features show some evidence for being
tightly wound spirals instead \citep[]{Cla03}. 
\item an evacuated cavity from 175 AU inwards \citep[]{Mou01, Cla03}, 
\item an arc-like feature between the inner and outer ring on the eastern 
side of the disc \citep[]{Mou01}, potentially part of the tightly
wound spiral claimed by \citet[]{Cla03}
\item a very faint open spiral structure on the 
edge of the disc which, towards the southwest, leads towards 
the binary companions HD141569BC \citep[]{Cla03}, potential evidence
for a fly-by event, and 
\item an offset of up to 30 AU between the star position 
and the measured disc centre \citep[]{Cla03}. 
\end{enumerate}

Numerous modeling efforts both with and without embedded planets 
have been attempted but none to date can capture
all the features of the imaged HD 141569A disc.
\citet[]{Tak01} model a mixed gas and dust ring and can reproduce an
inner and outer ring structure purely due to dust migration from gas friction
without invoking any embedded companions or interactions.  However, 
this modeling approach fails to reproduce any of the observed spiral
structure in this disc.  \citet[]{Aug04} and \citet[]{Qui05} model the
effects of the orbital passage of a bound companion (in this case,
presumably the binary M stars HD 141569BC) using an N-body 
gravitational code and a 2D hydrodynamical code respectively.
Such an interaction can reproduce the outer spiral features and faint
open spiral structure seen in
the HD 141569A disc, but fails to reproduce the inner spiral / ring or
inner cavity.  \citet[]{Wya05} models the effect of the secular
perturbations from a planet on an eccentric
orbit on the dust disc using an N-body gravitational code.  
This approach qualitatively reproduces the outer spiral feature, but
not the inner features, or the faint outer open spiral seen by
\citet[]{Cla03}.  Two more recent approaches combine the effects of a
fly-by event and one or more embedded planets \citep[]{Ard05, Rec09}, 
using N-body and SPH simulations.  These studies are generally able to 
reproduce the spiral structures seen with one planet and a fly-by, but
both struggle to reproduce the gap between the inner and outer spiral
/ ring structures.


We report here the tightest inner working angle (henceforth IWA) 
and highest near-IR resolution observations to date 
of the HD 141569A disc taken as part of the
NICI Science Campaign.  From 2008 December to 2012
September, the NICI Planet-Finding 
Campaign \citep[]{Liu10, Wah13a, Wah13b, Nie13, Bil13, Wah14, Nie14}
obtained deep, high-contrast AO imaging of $\sim$230 young, nearby
stars.  Our observations confirm much of the previously observed
structure, reveal new asymmetries in this disc, and place strong
upper limits on the mass of any embedded planet.

\section{Observations and Data Reduction}

The near-IR coronagraphic imager (NICI) was the first custom planet-finding camera, available at the
Gemini South 8-m telescope from 2008-2012. 
NICI is a dedicated adaptive optics (AO) instrument tailored expressly for direct
imaging of exoplanet companions (Chun et al. 2008), combining several techniques to attenuate starlight and suppress
speckles for direct detection of faint companions to bright stars: (1)
Lyot coronagraphy, (2) dual-channel
imaging for Spectral Differential Imaging (SDI; Racine et al. 1999; Marois et al. 2005;
Biller et al. 2007), and (3) operation in a fixed Cassegrain rotator mode for Angular Differential
Imaging (ADI; Liu 2004; Marois et al. 2006; Lafreni\`ere et al. 2007).
As part of the NICI 
Planet-finding Campaign, HD 141569 was observed with NICI on
2009-03-07, 2010-04-08, and 2011-05-03.  Observations and
sky rotation over the observation are reported in Table~\ref{tab:obs}.
Observations were conducted in the two standard NICI campaign modes:
Angular Spectral Differential Imaging (ASDI) and Angular Differential
Imaging (ADI).  In both modes, the telescope rotator was left off,
allowing the field to rotate with parallactic angle on the sky, 
enabling effective speckle suppression.  In 
ASDI mode, we observed simultaneously in two narrow band filters
(1.578 $\mu$m; $CH_{4}S$ 4\% and 1.652 $\mu$m; $CH_{4}L$) which cover
the 1.6 $\mu$m absorption feature found in T-type brown dwarfs.
To prevent saturation and yield higher contrasts, the primary star was
always placed behind NICI's partially transmissive focal plane mask.
Frame time was kept to 60 s / frame across the NICI campaign, with
each frame consisting of multiple coadded subframes to prevent saturation.
We reduced each dataset using two independent pipelines: 1) a 
custom ADI pipeline developed specifically for the NICI Campaign \citep[]{Wah13b} 
and 2) the publically available PCA (principal component analysis)
pipeline of Dimitri Mawet (http://www.astro.caltech.edu/$\sim$dmawet/pca-pipeline.html).

The NICI Campaign pipeline is described in
detail in \citet[]{Wah13b}.  We describe the basic data reduction only
very briefly here.  After basic flatfield and distortion correction, each
image is registered to a master image based on the centroid position
of the primary star under the mask.  A PSF image is built from the
median combination of the entire stack of images and then subtracted
from each individual image.  The individual images are then rotated
to place north up and east to the left and median combined.  
For the purposes of this study, data from each channel taken in ASDI
mode were reduced separately.

For the PCA reduction, 
basic data processing was done similarly as with the NICI Campaign pipeline.
PCA-based pipelines (following the algorithms of
Soummer et al. 2012, Amara \& Quanz 2012) use a stack of images 
(taken at different times or different wavelengths) to determine the 
principle components of the data.  These principal components are then 
used to build ideal PSFs image by image for quasi-static speckle
suppression.  For each NICI dataset, we reduced data from each 
narrowband filter separately, using 
the individual frames from a single narrowband filter over 
the ADI time sequence to generate
the principal components and build ideal PSFs.
After ideal PSFs are built and
subtracted, each frame is then derotated to place north up and east to
the left.  Finally, all frames are stacked to provide the final image.

We retrieve the HD 141569 disc in all 4 NICI datasets.
However, the 2009 and 2010 datasets rotate by $<$30$^{\circ}$ on the
sky and thus, the disc is more affected by self-subtraction and
is comparatively faint in these datasets.  Therefore, we only present results from the
2011-05-03 dataset in this paper.

\section{Results}

\subsection{Pipeline results}

\begin{figure}
\includegraphics[width=3in]{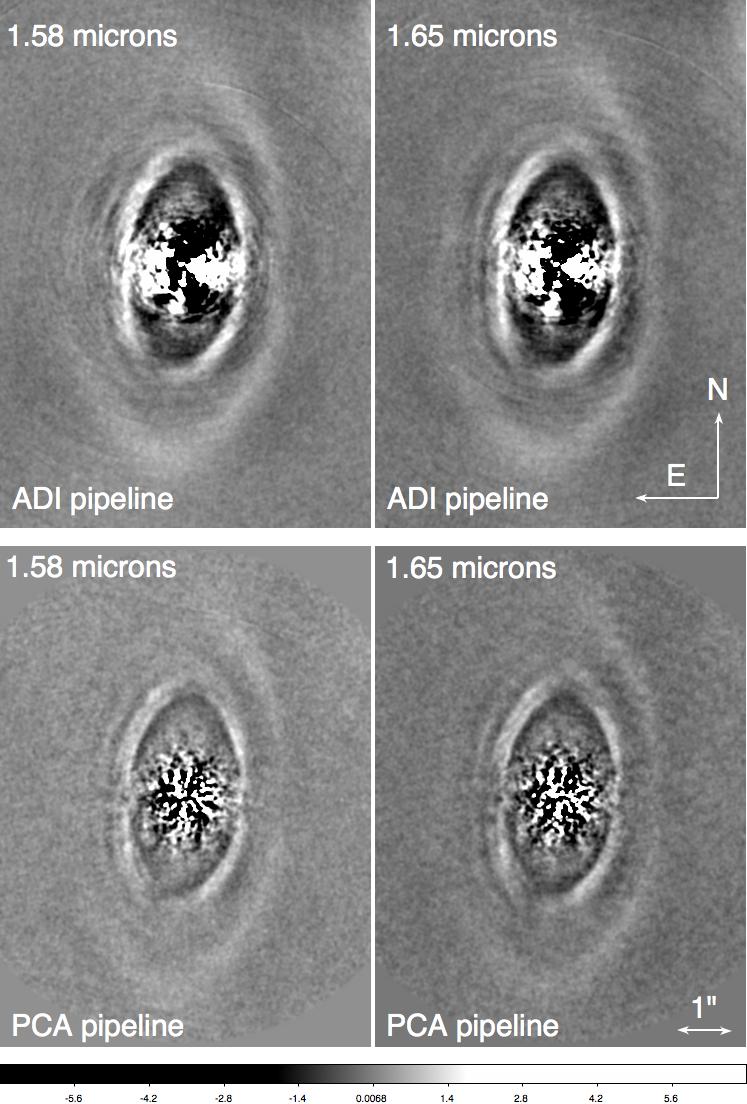}
 \caption{Reduced NICI images of the HD 141569 disc.  Images have been
 smoothed with a 3-pixel Gaussian to highlight faint structures. \label{fig:images}}
\end{figure}

\begin{figure}
 \includegraphics[width=3in]{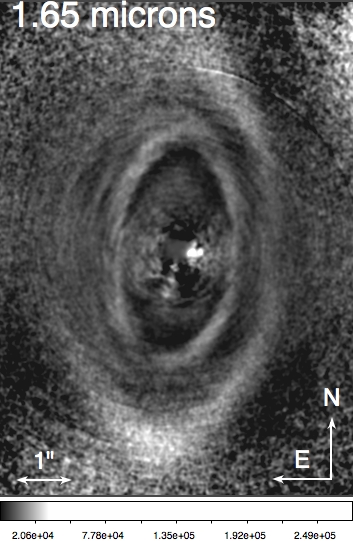}
 \caption{Reduced NICI images of the HD 141569 disc.  Images have been
 smoothed with a 3-pixel Gaussian and scaled by $r^{2}$ (radial
 separation from the image center) to highlight faint structures at
 wide separations. \label{fig:imscale}}
\end{figure}

\begin{figure}
 \includegraphics[width=3in]{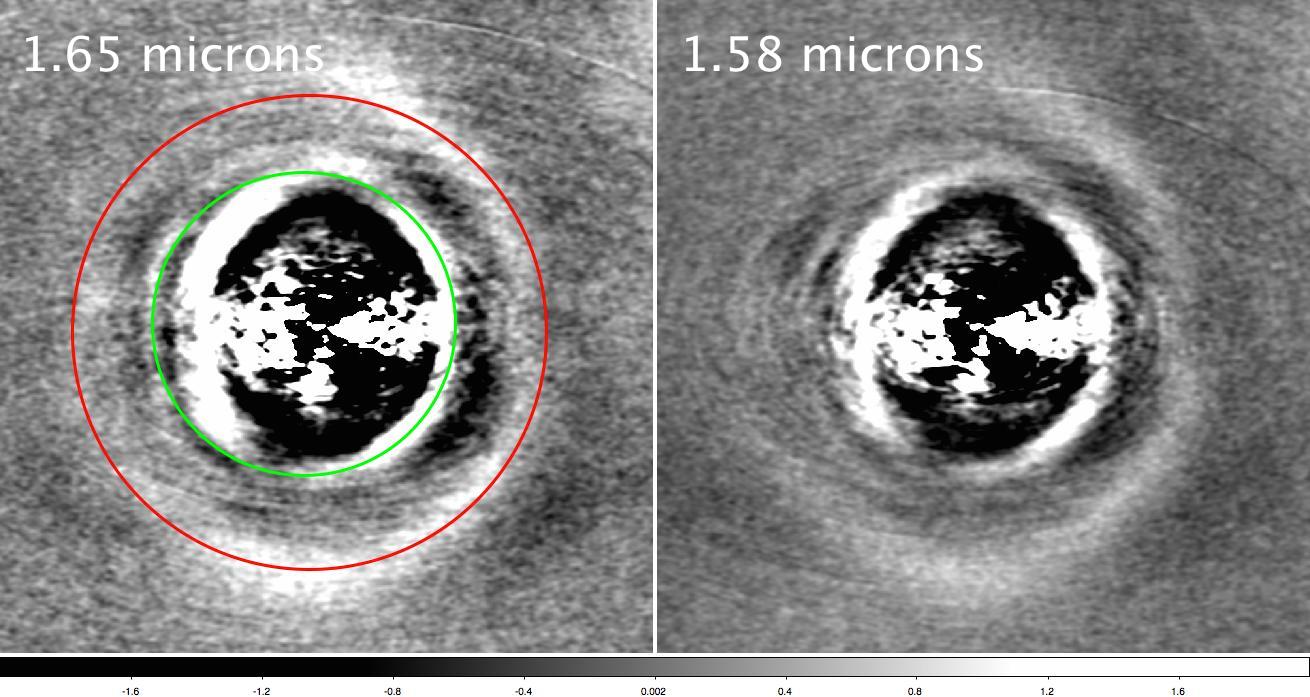}
 \caption{Deprojected disc images. 
While the outer ring structure looks circular in this deprojection,
the inner bright ring looks rather elliptical.  This suggests that a
single deprojection angle is not appropriate for this system
and that there may be an offset in PA between the two ring / spiral
features.\label{fig:deproj}}
\end{figure}

Reduced images from both pipelines and both NICI channels 
are presented in Fig.~\ref{fig:images}.  Images have
been smoothed with a 3-pixel Gaussian to highlight faint structures.
Disc features in the ADI and PCA-reduced images appear roughly similar, 
although it is clear that the disc self-subtraction is worse in 
in the PCA reduction.  In order to highlight faint outer structure, 
the 1.65$\mu$m ADI-processed
 image scaled by $r^{2}$ (radial separation from the image
center) is presented in Fig.~\ref{fig:imscale}
An annotated image with main features labeled is presented in 
Fig.~\ref{fig:annot}.
A deprojected image is presented in Fig.~\ref{fig:deproj}, using an
inclination to the line of sight of 51$^{\circ}$ (where 90$^{\circ}$
is edge-on and 0$^{\circ}$ is pole-on) and a PA of
356$^{\circ}$ (from \citet[]{Wei99}).

\subsection{Features recovered}

We recover 5 main features in the NICI images of the HD 141569 disc
discovered in previous HST imaging, however, 
significant self-subtraction is apparent in both our ADI and PCA
reductions (Fig.~\ref{fig:images}) which affects the apparent
morphology of some of these features:

1) an inner ring / spiral feature \citep[]{Cla03,Mou01, Wei99}. 
In the NICI imaging, self-subtraction will affect the details of ring or
spiral feaures, so we cannot confirm nor refute the claim 
from \citet[]{Cla03} of tightly wound spirals.
 Once deprojected, this feature does not appear circular.

2) an outer ring / spiral \citep[]{Cla03,Mou01, Wei99, Aug99} which 
is considerably brighter on the western side compared to the eastern
side, but looks fairly circular in the deprojected image.

3) an additional arc-like feature between the inner and outer ring
only evident on the east side of the image and only imaged in the IR before by 
\citet[]{Mou01}.  This may correspond to the tightly wound spiral
claimed by \citet[]{Cla03}.  In the deprojected image,
this feature appears to complete the circle of the west side inner ring.

4) a lower surface brightness region between the inner and outer 
ring/spiral features.

5) an evacuated cavity from 175 AU inwards \citep[]{Mou01, Cla03}, 

We do not retrieve the wide ``open spiral arms'' found by
\citet[]{Cla03}, but we may simply not be sensitive to such 
low surface brightness features; working
in the near-IR from the ground, we are limited by a much higher sky
background compared to the optical HST-ACS imaging of \citet[]{Cla03}.

\begin{figure}
 \includegraphics[width=3in]{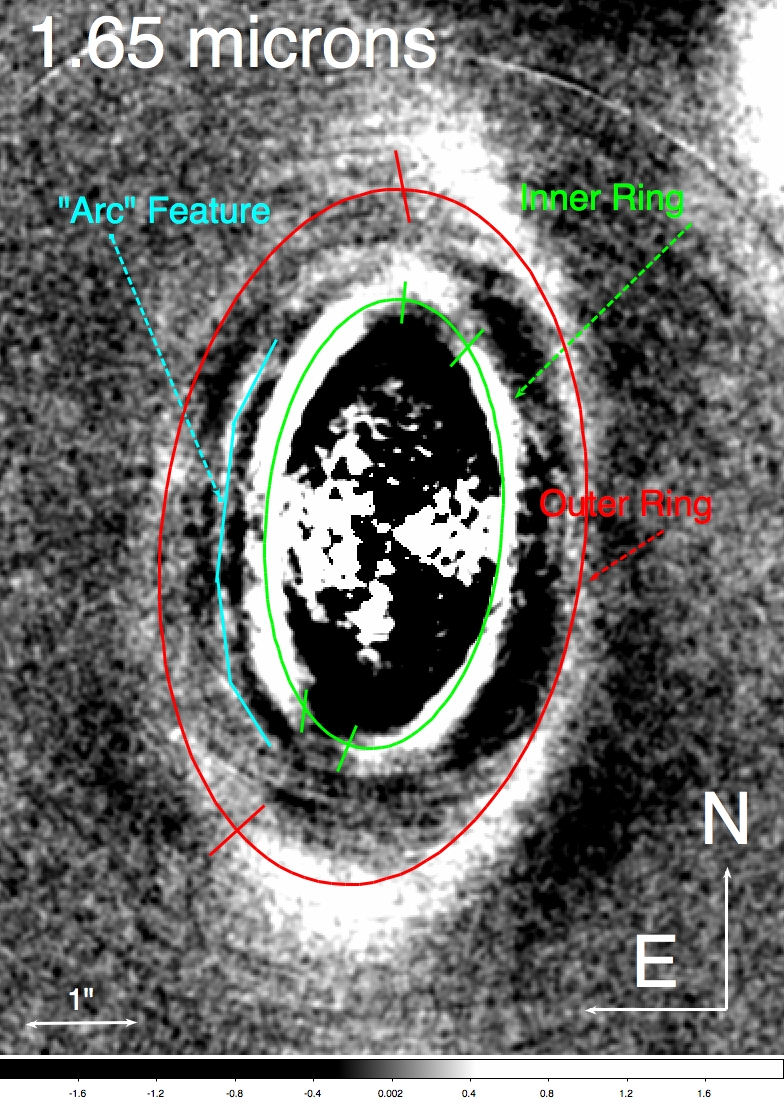}
 \caption{Reduced NICI image of the HD 141569 disc with key features
   annotated, including: 1) an inner ring / spiral feature
   \citep[]{Cla03,Mou01, Wei99},  
2) an outer ring \citep[]{Cla03,Mou01, Wei99, Aug99} which 
is considerably brighter on the western side compared to the eastern
side. 
3) an additional arc-like feature between the inner and outer ring
only evident on the east side of the image and only imaged before by 
\citet[]{Mou01}, and 
4) an evacuated cavity from 175 AU inwards \citep[]{Mou01, Cla03},  
 Images have been
 smoothed with a 3-pixel Gaussian to highlight faint structures.\label{fig:annot}}
\end{figure}

We note some additional asymmetries in this system.  Specifically, 
while the outer ring / spiral structure appears circular in this deprojection,
the inner bright ring / spiral appears elliptical.  This suggests that a
single deprojection angle is not appropriate for this system
and that there may be an offset in inclination angle between the two ring / spiral
features.  Additionally, there appear to be gaps to the north and
south in the inner ring / spiral feature (Fig.~\ref{fig:annot}), which
may be artifacts from disc self-subtraction.

\subsection{Assessing Self-Subtraction}

Significant self-subtraction is apparent in both our ADI and PCA
reductions (Fig.~\ref{fig:images}), and appears somewhat worse in the
PCA reduction.  
Over the course of the observation, the disc rotates $\sim$60 degrees on the sky.  
Given the extent of the disc, no matter how carefully we build our ADI
PSF, the final image will still suffer from self-subtraction.  
Unlike previous HST observations which used reference star subtraction
to remove the stellar PSF and reveal the disc,
self-subtraction is unavoidable for circumstellar discs with all
ADI-based observations but can be calibrated in some cases with
careful data reduction.  Completely symmetric face-on
discs will be entirely subtracted using any ADI-based algorithms, 
whereas edge-on discs will be minimally affected, and only in their
inner regions. For edge on discs, conservative ADI algorithms 
can prevent most self-subtraction, however, 
with an inclination of 51$^{\circ}$, the HD 141569 disc is 
right on the edge of being amenable to ADI reduction.  
The complex double ring / spiral structure of this disc
will also introduce further complications in this case.

Milli et al. 2012 directly consider the case of ADI with the 
HD 141569 disc.  They show
that a classical ADI reduction will act like a filtering algorithm,
with the following effects: narrowing of disc features along the minor
axis and brightening at the ansae (as seen e.g. in HR 4796, 
Thalmann et al. 2011).  Generally for discs with 
various morphologies, they find that while ADI will affect
the flux brightness along the disc, 
measurements of disc centre position, disc PA, and disc inclination
are minimally affected.


Given these considerations, we only qualitatively describe disc
morphology in this paper, attempt to identify and point out potential
artifacts, and do not measure flux in the disc.  We first fit a simple
double ring model to the reduced data as a baseline estimate of the 
disc morphology.  
We then introduce the double ring model as well as a spiral model 
into a separate NICI dataset with similar field rotation
to determine if we can distinguish between these two scenarios.

\subsubsection{Ring Parameter Fitting}

As a baseline quantification of the morphology of the disc structure,
we fit each apparent ring feature
using a Markov Chain Monte Carlo (MCMC) method similar to that used by \citet[]{Wah14} for the HR
4796A disc.  As we find no significant differences between the imaged morphology in
the 1.65 $\mu$m image vs. the 1.58 $\mu$m image, we conduct the MCMC
fitting only on the 1.65 $\mu$m image.
This model is intended solely as a baseline to understand how self-subtraction
may have affected disc structure and to measure some basic
morphological properties of the disc.

We model the geometry of each feature as a a circular annulus with mean
radius r$_{0}$ and a gaussian radial brightness profile with width
$\sigma$:

\begin{equation}
B(r) = B_{0} e^{-(r_{0} - r)^{2}/2\sigma^{2}}
\end{equation}

with additional parameters including the disc inclination, disc PA, 
and position of the ring center, for 7 parameters total. 
While significant azimuthal asymmetries have been found by several
authors \citep[][]{Wei99, Mou01, Cla03}, the details of these
asymmetries are too complicated to model analytically and will also be
affected by our ADI subtraction.  Thus we have
chosen this simple model for the moment and accept the fact that the 
brightness fit for each ring will be incorrect. 

We use the AMOEBA algorithm to find 
an initial best-fit set of parameters.  We then use
Metropolis-Hastings Markov Chain Monte Carlo methods to evaluate the
posterior probabilities for the 7 ring parameters.

Starting from the initial parameter set, we apply a trial
jump to a new part of parameter space.  The trial jump is 
drawn randomly from a seven-dimensional Gaussian distribution across 
all 7 ring parameters with appropriate standard deviations along each
dimension.  We evaluate the goodness of fit of the trial model using the
$\chi^{2}$ statistic:

\begin{equation}    
\chi^{2}  = \Sigma_{pixels} \frac{(Data - Model)^{2}}{noise^{2}}
\end{equation}

The trial jump is accepted or not according to the Metropolis-Hastings
measure and the algorithm is run for 10$^{6}$ steps. For a more in-depth discussion of 
this method, see \citet[]{Wah14}.

\begin{figure}
 \begin{tabular}{cccc}
\includegraphics[width=1.5in]{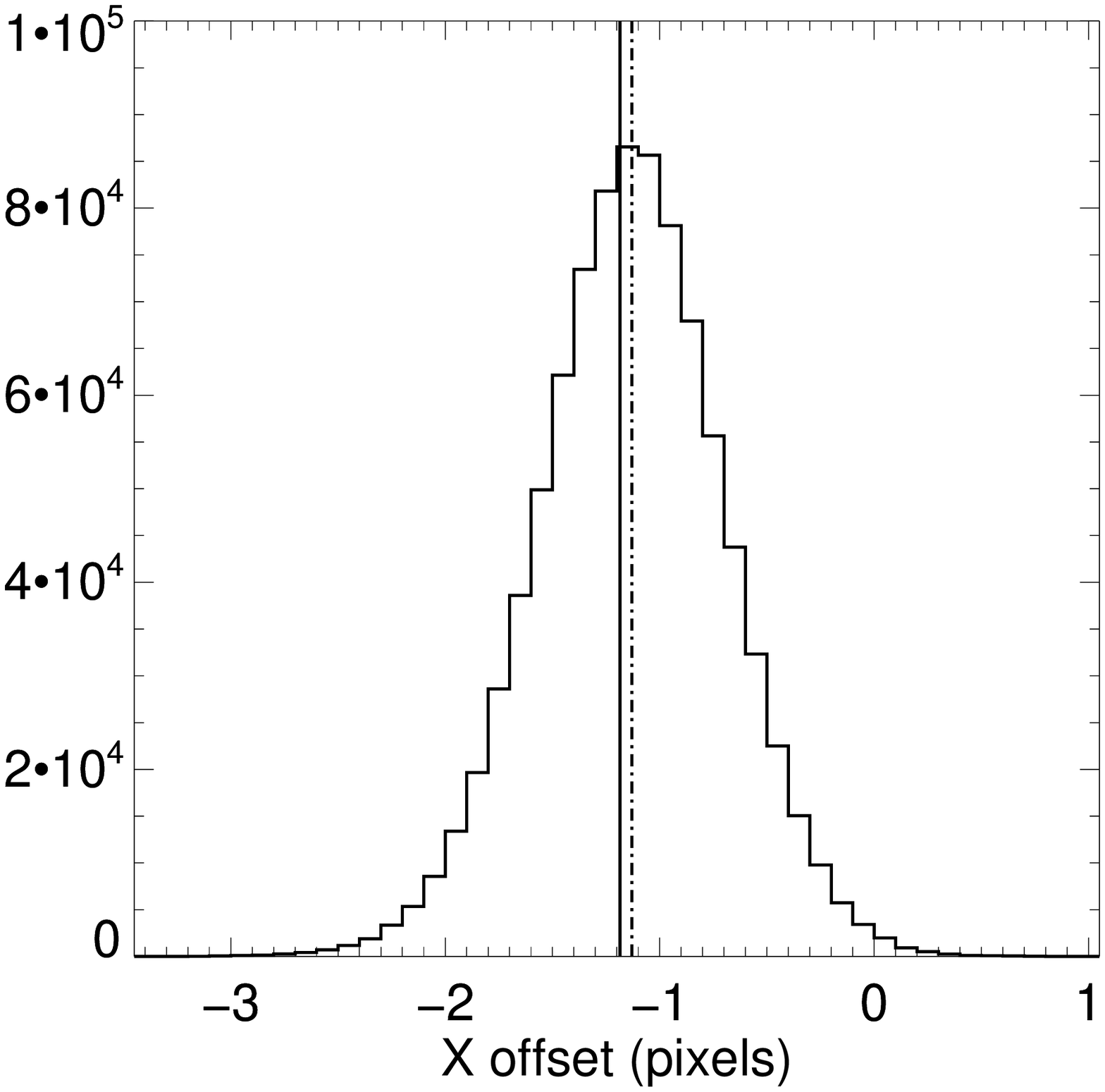} & \includegraphics[width=1.5in]{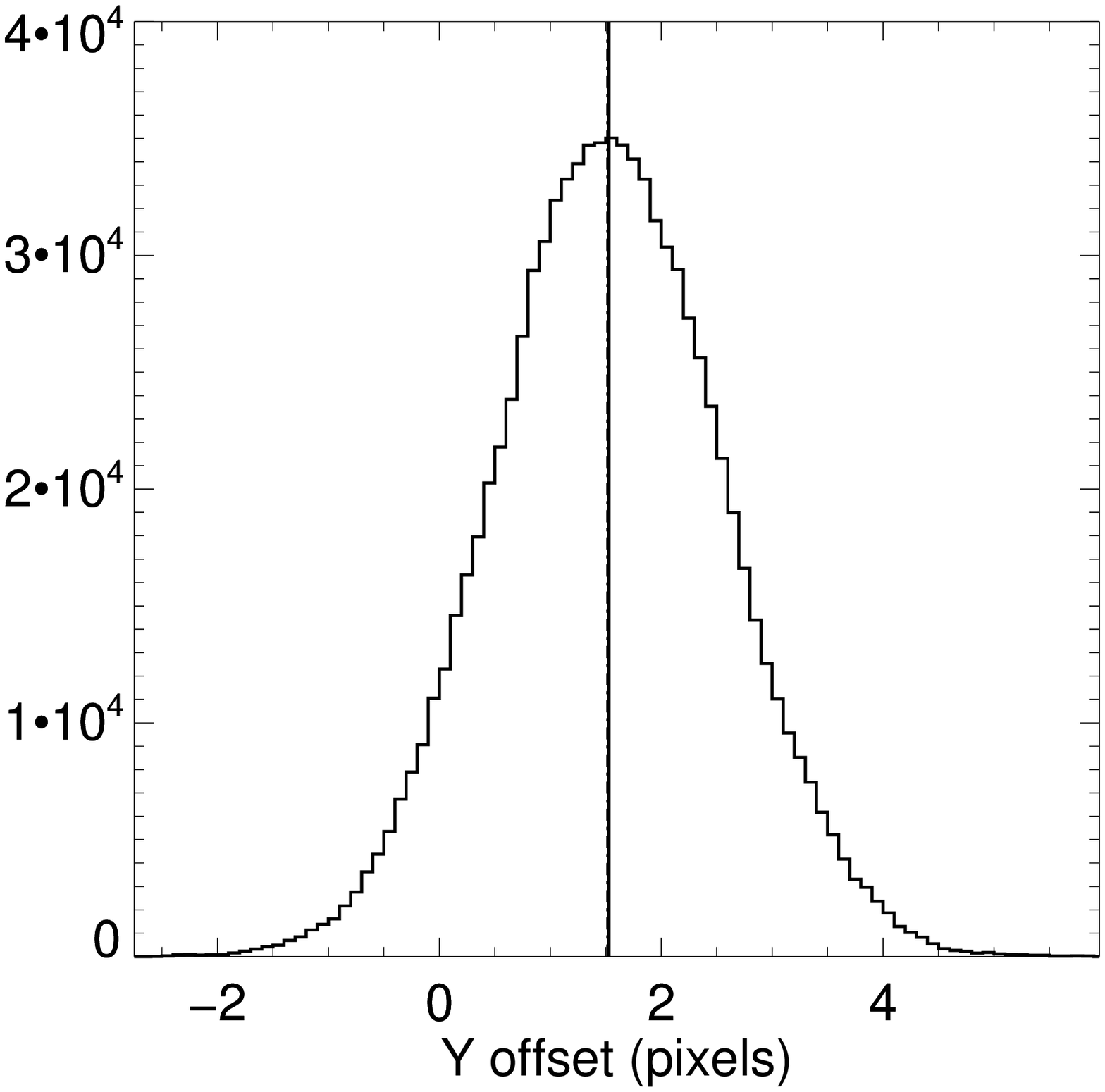}
& \includegraphics[width=1.5in]{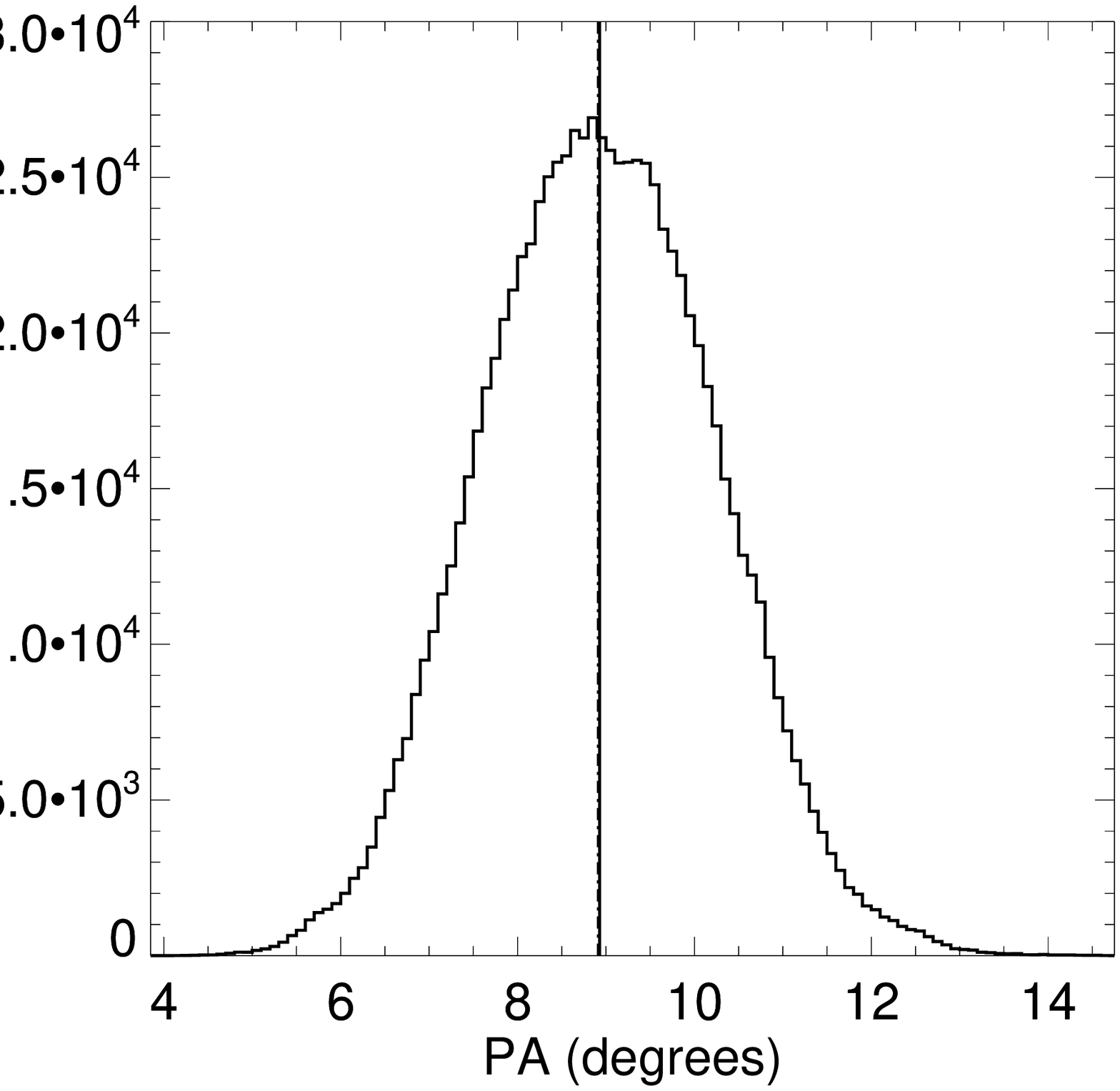}
& \includegraphics[width=1.5in]{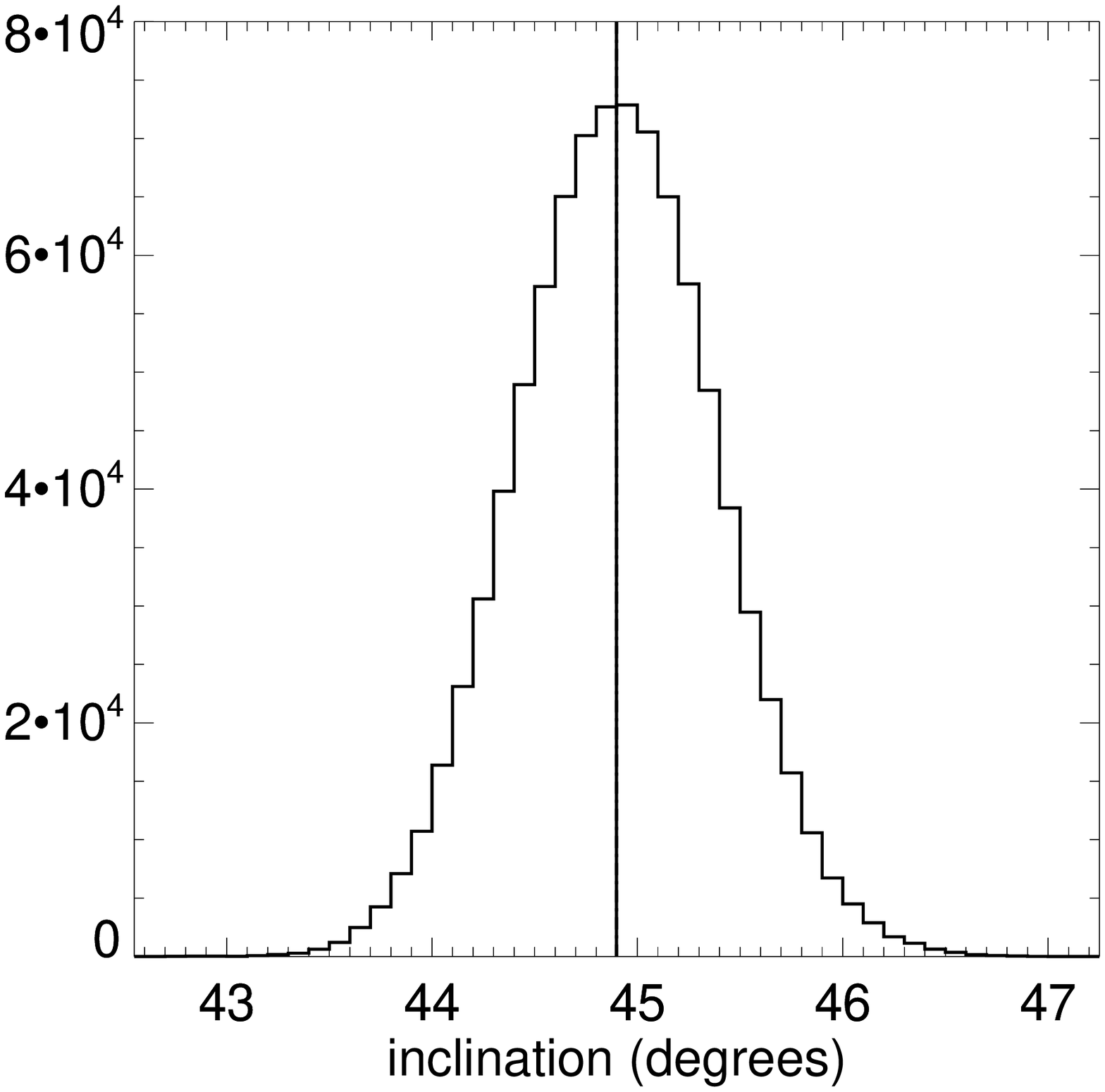} \\
\includegraphics[width=1.5in]{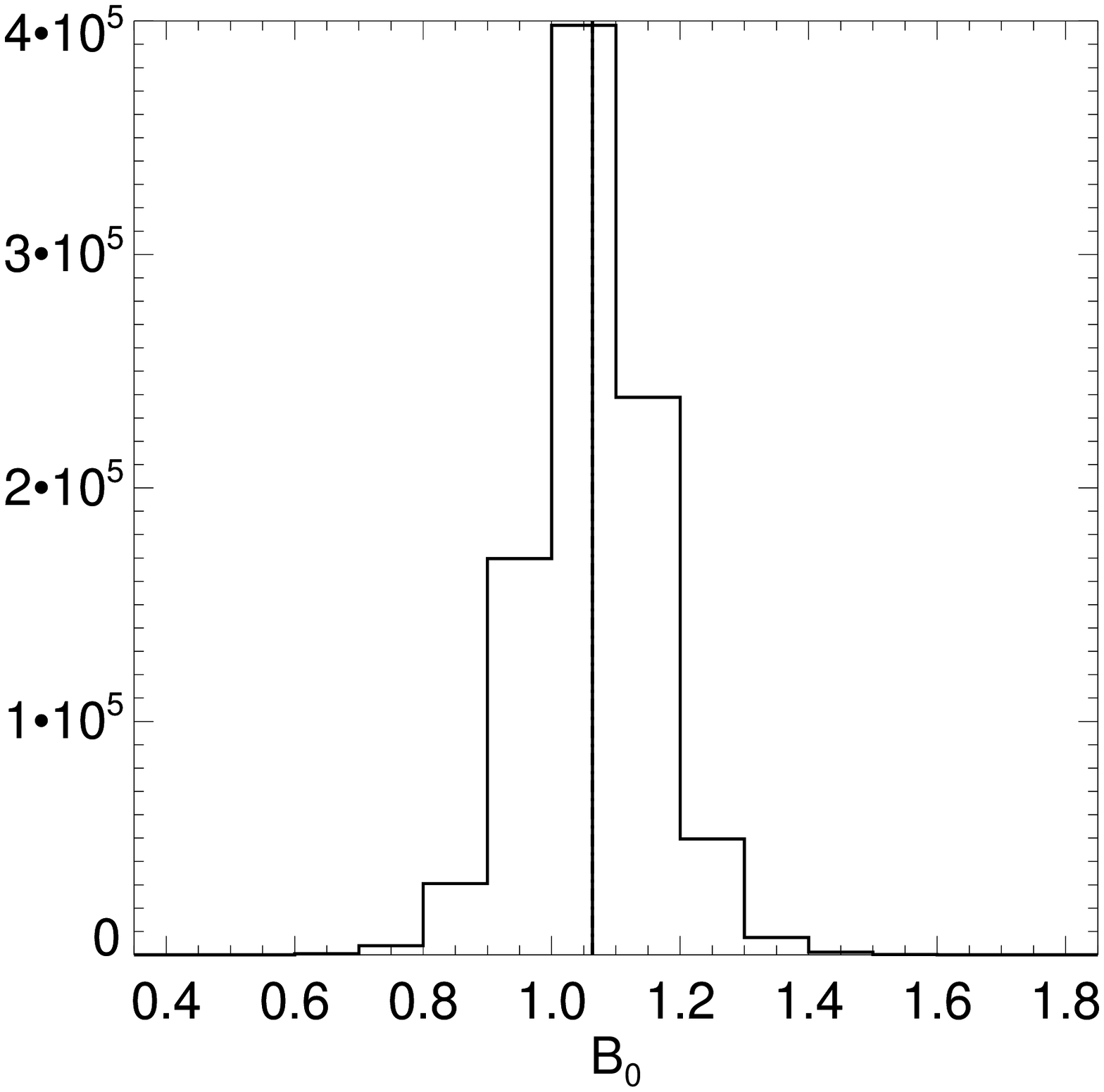} & \includegraphics[width=1.5in]{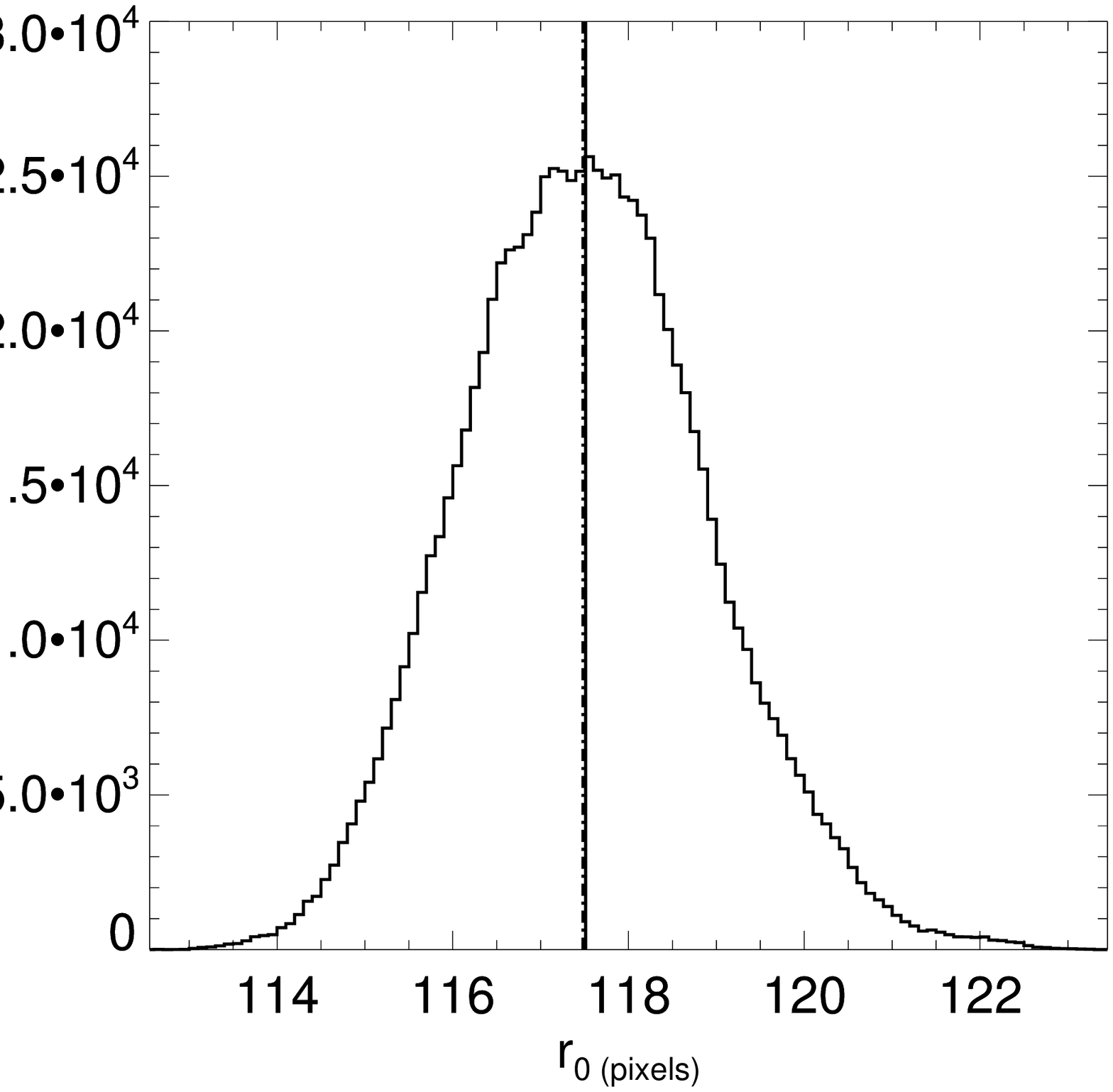}
& \includegraphics[width=1.5in]{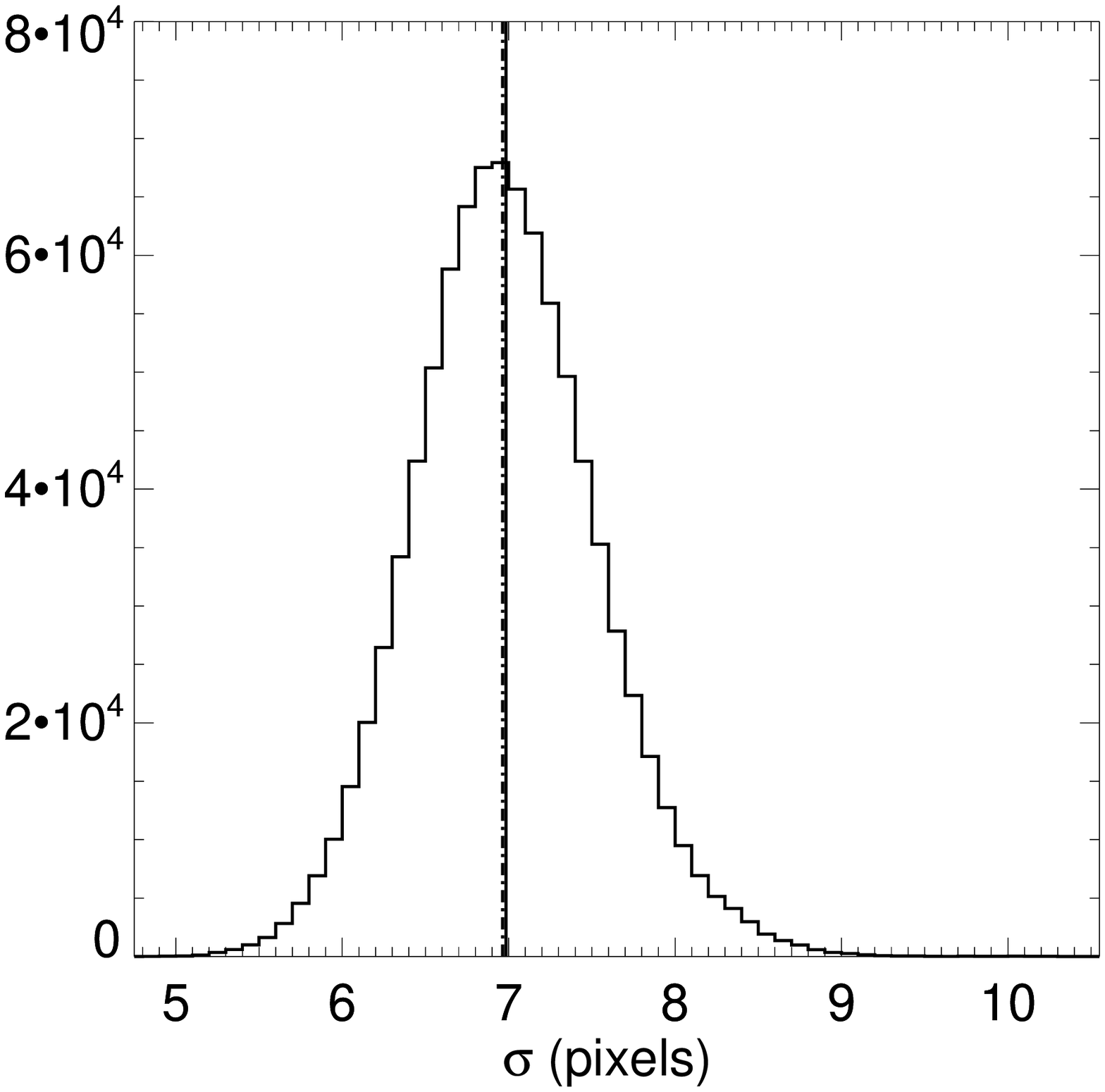}
&  \\
\end{tabular}
\caption{Posterior probability distribution functions for the ``inner ring''\label{fig:pdfsinner}}
\end{figure}

\begin{figure}
 \begin{tabular}{cccc}
\includegraphics[width=1.5in]{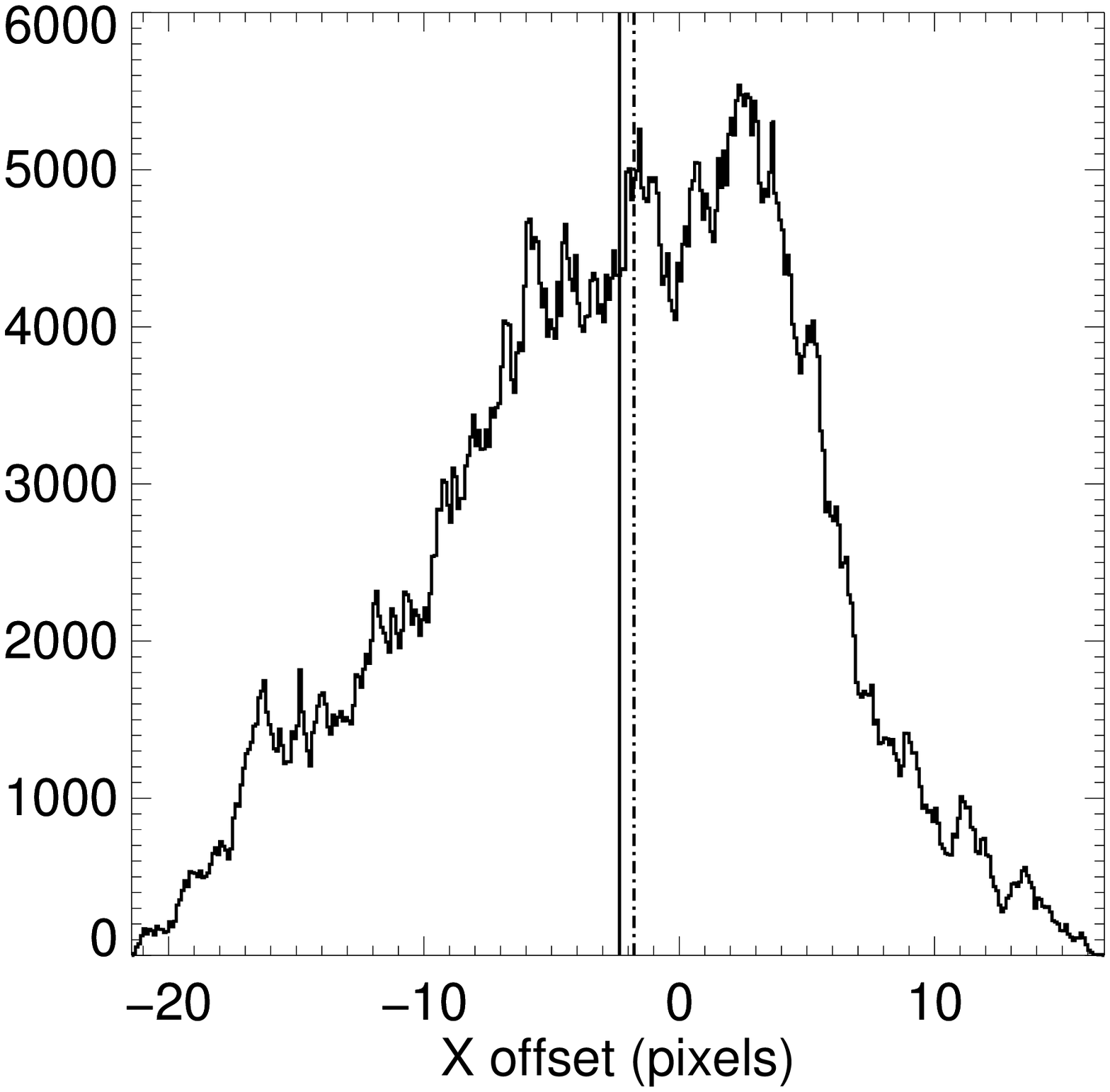} & \includegraphics[width=1.5in]{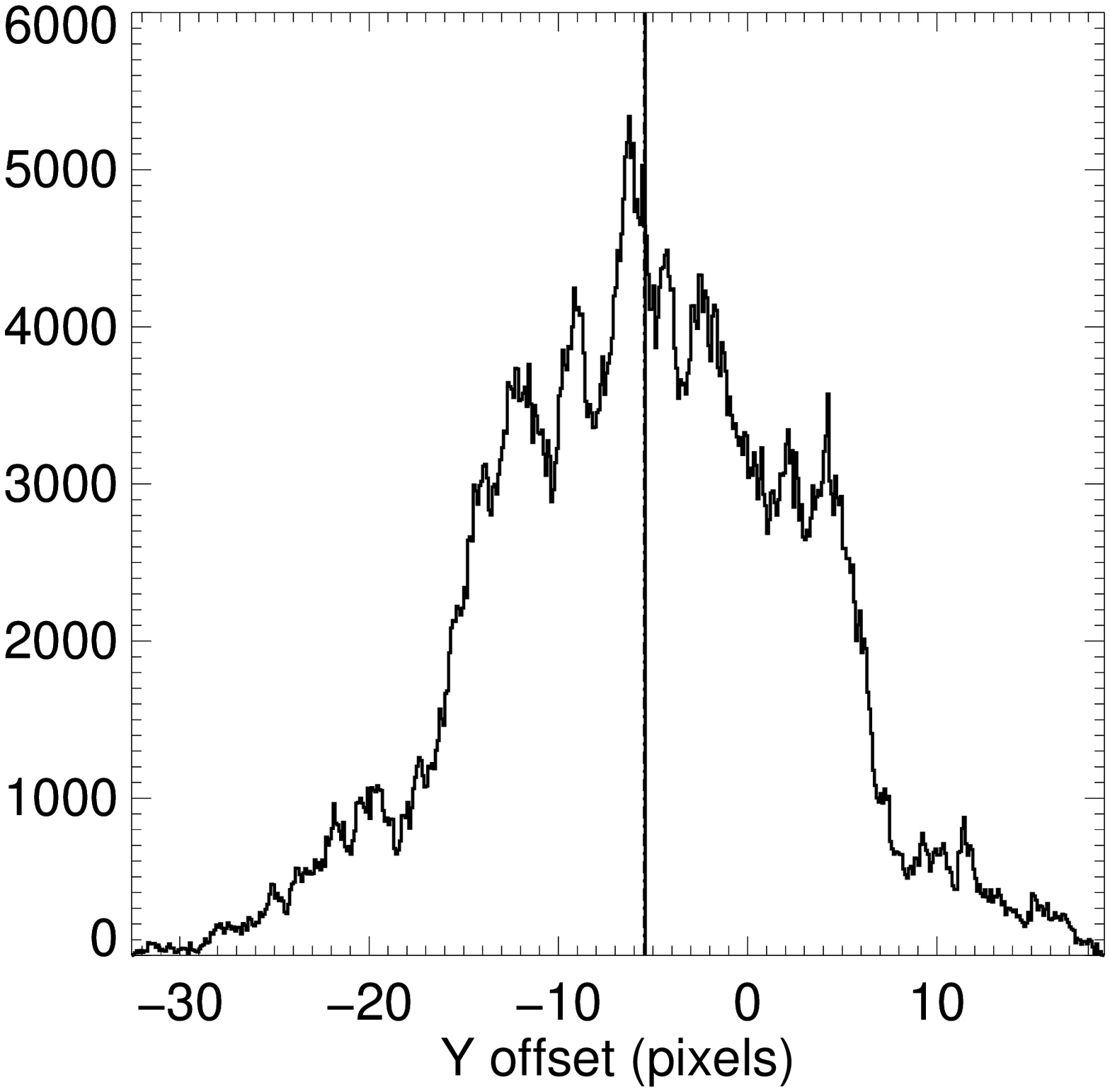}
& \includegraphics[width=1.5in]{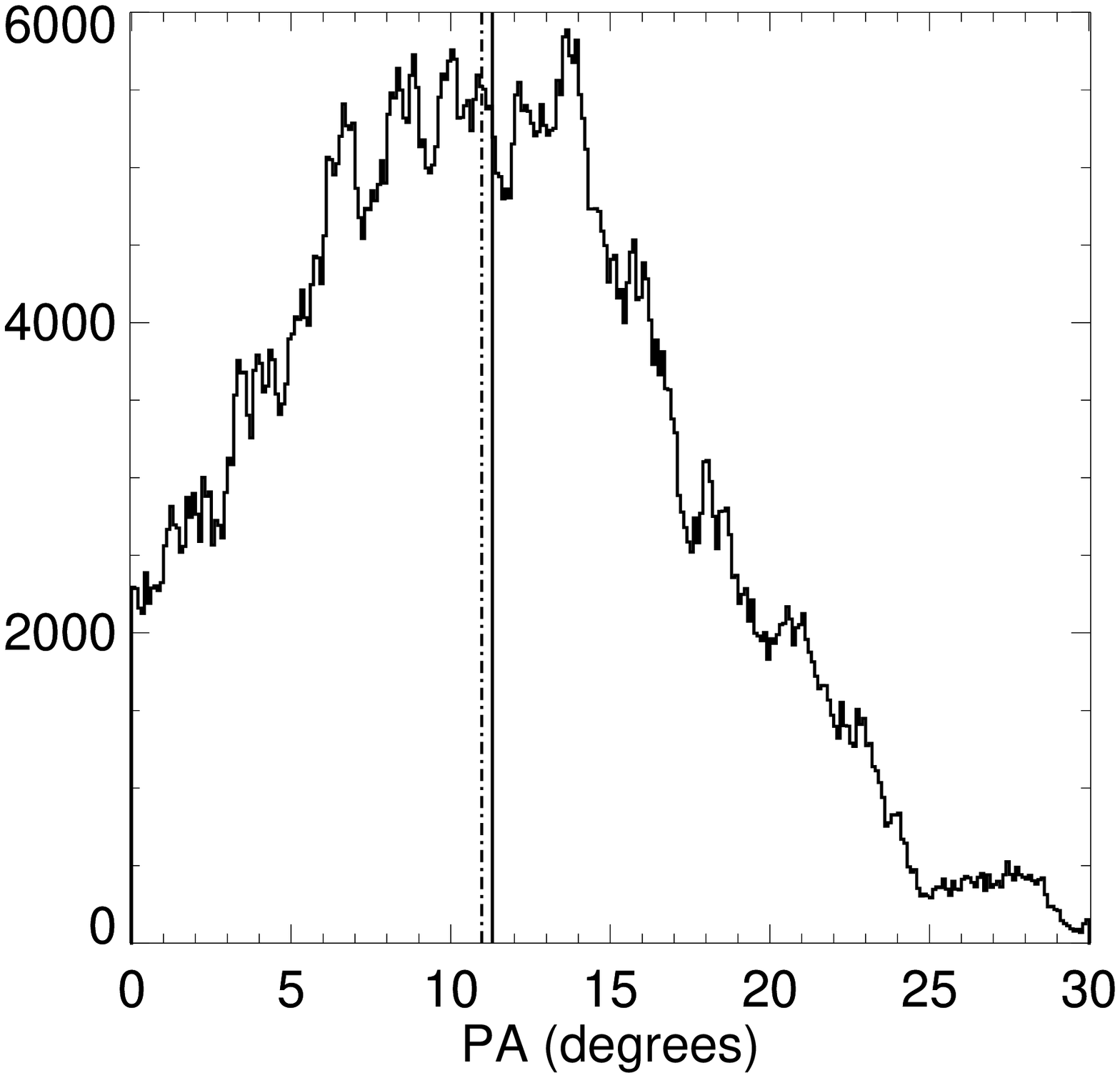}
& \includegraphics[width=1.5in]{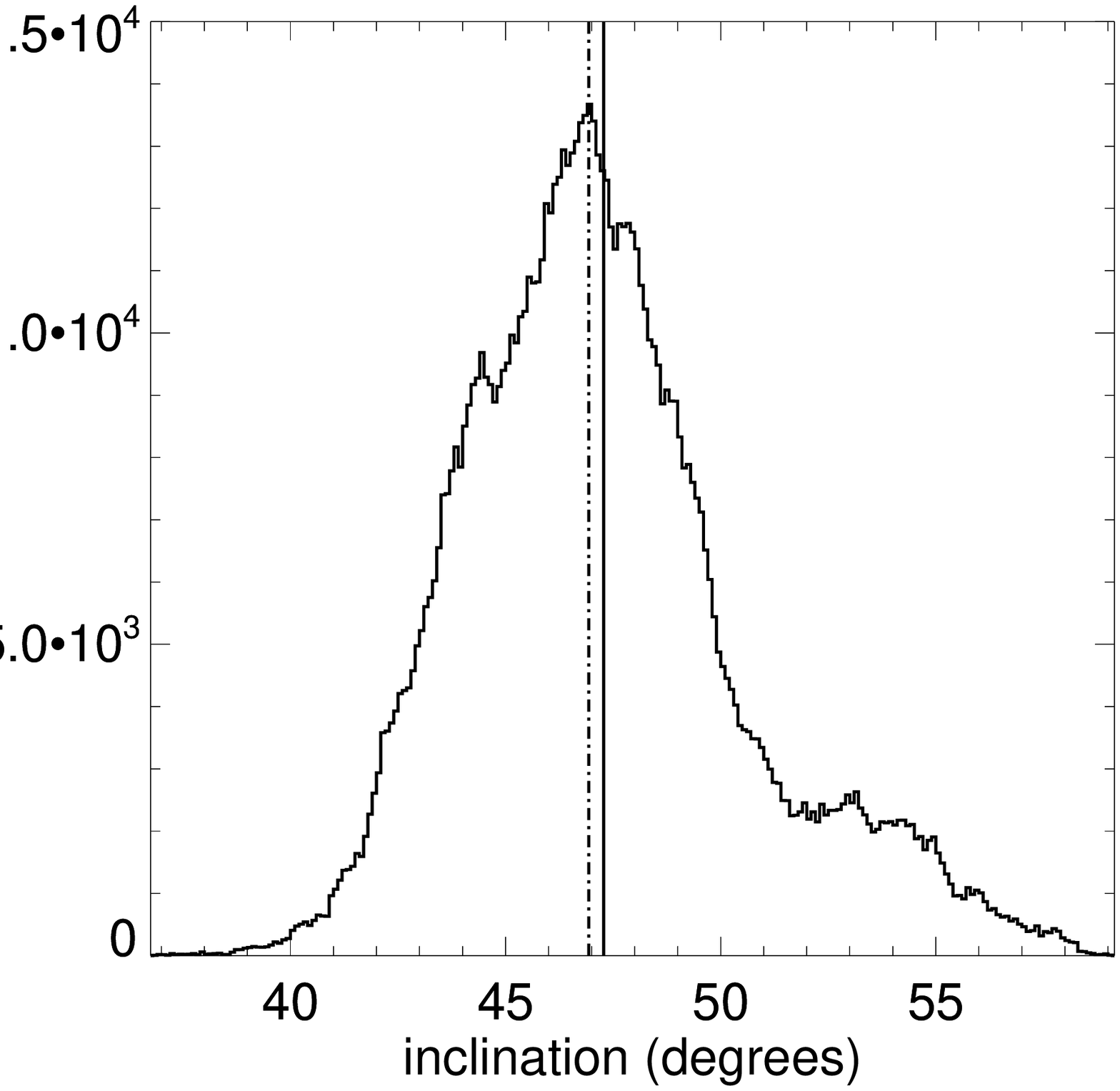} \\
\includegraphics[width=1.5in]{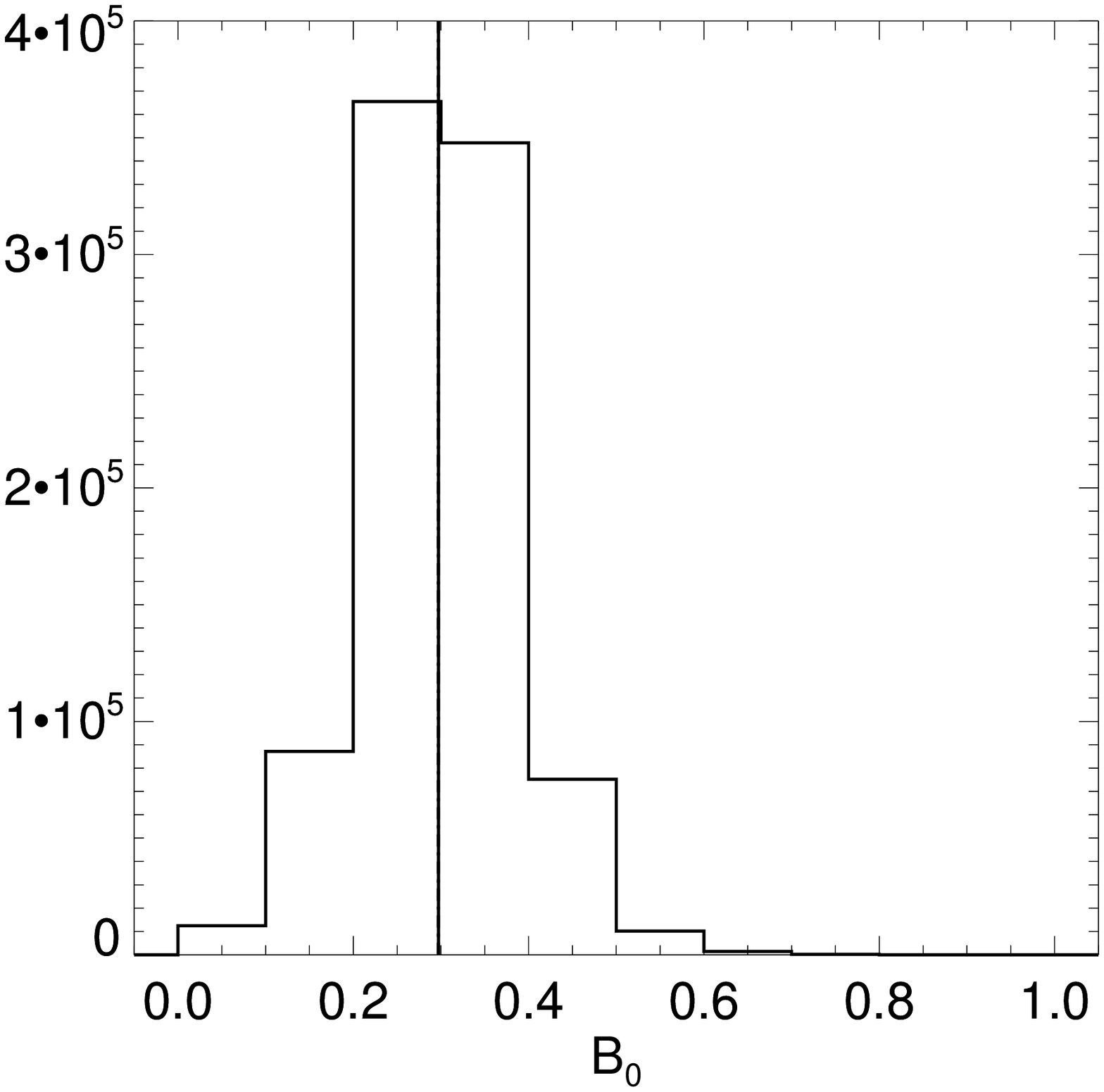} & \includegraphics[width=1.5in]{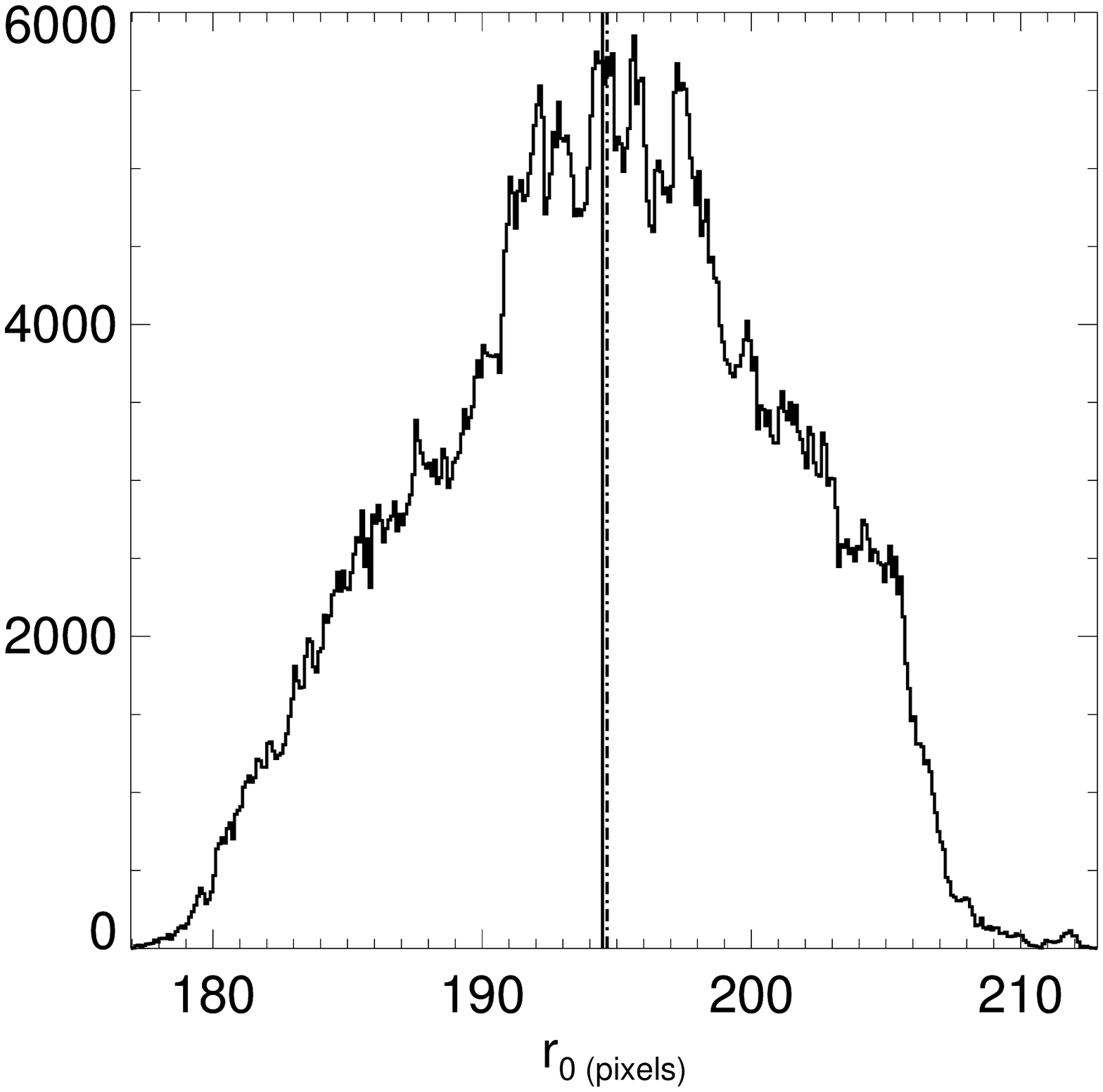}
& \includegraphics[width=1.5in]{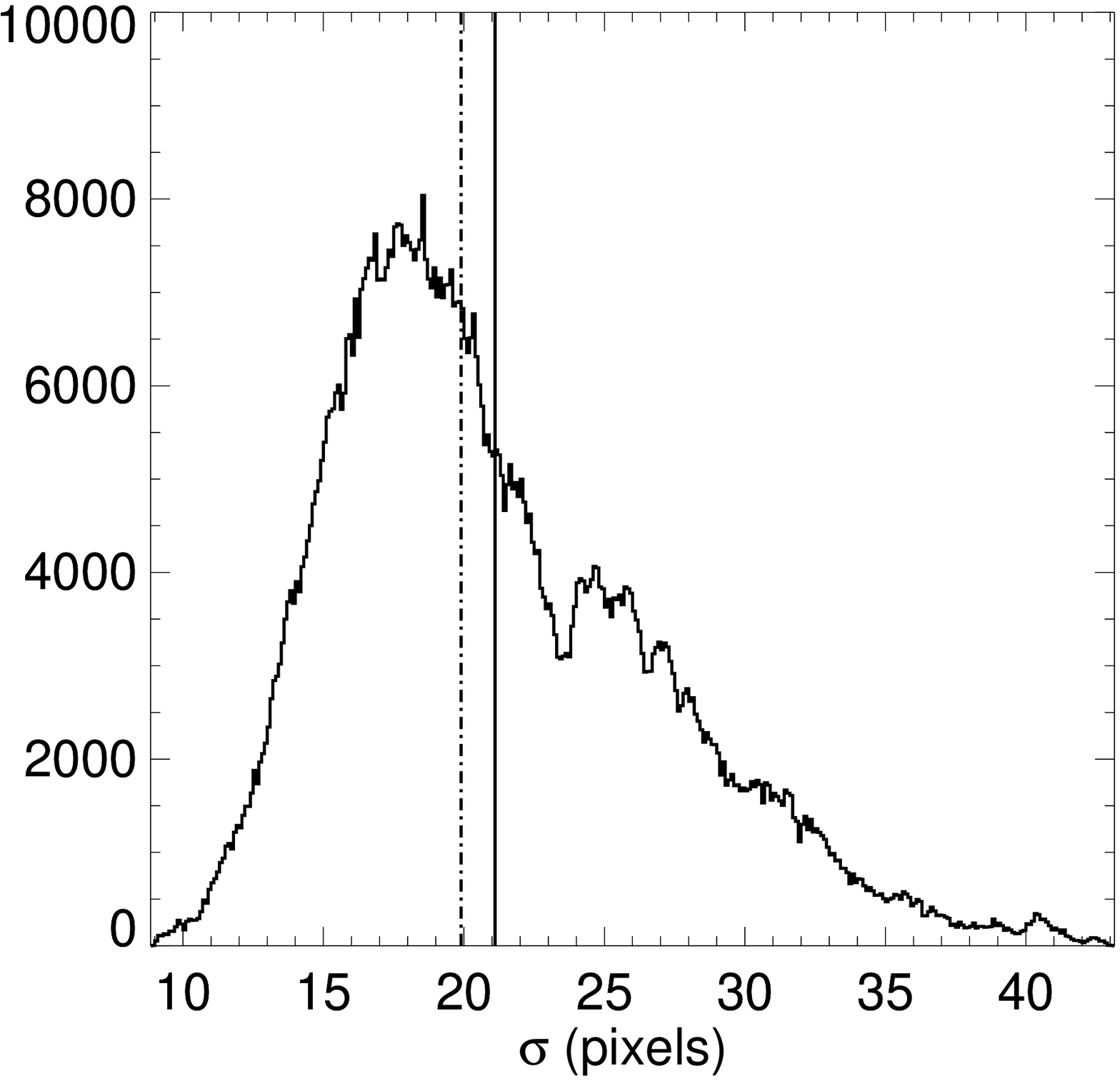}
&  \\
\end{tabular}
\caption{Posterior probability distribution functions for the ``outer ring''\label{fig:pdfsouter}}
\end{figure}

\begin{figure}
\includegraphics[width=5in]{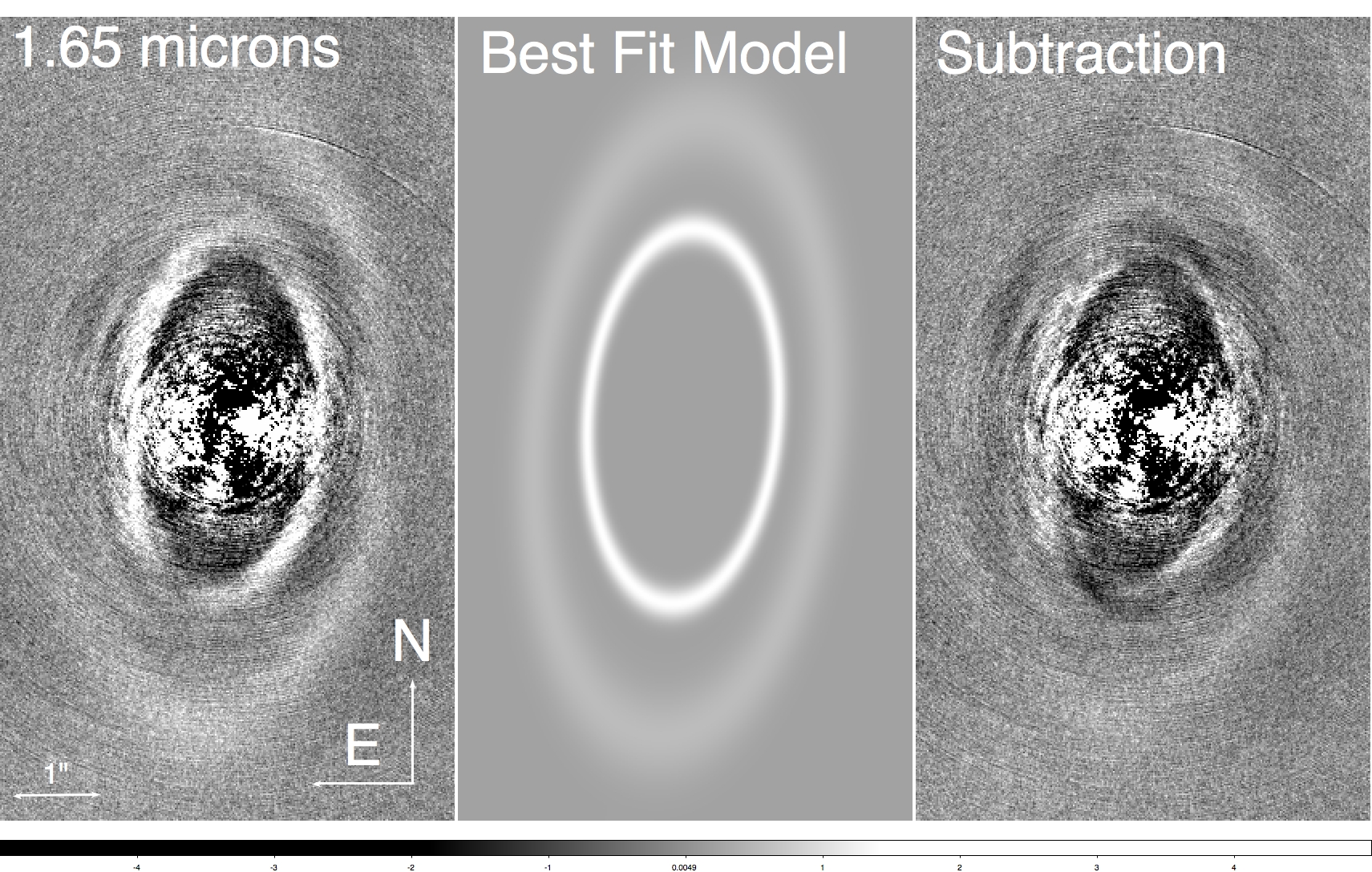}
\caption{Model subtracted images.  Left: reduced disc image, center:
best model image (median of the 7-parameter posterior pdf), right:
subtraction of best model image from the reduced disc image.
The Gaussian ring model used does not account for the clearly observed azimuthal
brightness asymmetries.
\label{fig:subs}}
\end{figure}

We evaluated the posterior probability distribution functions (henceforth
pdfs) for both the inner and outer ring
features.  Chains were visually inspected to ensure convergence.
We also used the Gelman-Rubin (GR) statistic to check for convergence.
After throwing out the first 10$^{5}$ steps to account for burn-in, we
divided each chain into three 3$\times$10$^{5}$ step sub-chains.
To calculate the GR statistic, we calculate the variance in each
parameter in each sub-chain as well as the total variance across all chains.
The GR statistic is then the ratio of the variance in each parameter in individual sub-chains
to the total variance for all sub-chains.  If the sub-chains are
exploring the same (converged) part of parameter space, the GR
statistic should be close to 1 and values $\sim$1.1 and below indicate
convergence.  By 3$\times$10$^{5}$ steps, all parameters have converged 
for both rings except for B$_{0}$, the peak brightness value of the rings.
 MCMC posterior pdfs are presented in Fig.~\ref{fig:pdfsinner} and ~\ref{fig:pdfsouter}  and best
 parameter values, uncertainties, and GR statistic values are
 presented in Table~\ref{tab:params}.  A subtraction of the ``best''
 models for both inner and outer rings (defined here as the median model from the evaluated posterior
 pdfs) is presented in Fig.~\ref{fig:subs}.

We can glean several important results from the MCMC fits.  First, we
recover an offset between the star and the disc center for both rings,
similar to \citet[]{Cla03}.  However, we find that the inner ring is
offset by 4$\pm$2 AU from the star center, 
compared to up to 30 AU found by \citet[]{Cla03}.  The
offset for the outer ring is larger ($\sim$12 AU), but with
considerably larger error bars.
Nonetheless, these measured offsets point to possible perturbing
companions in this disc.  The fact that B$_{0}$ does not converge is 
unsurprising -- our model assumes uniform azimuthal brightness of 
both rings, and this is clearly not the case.  Thus, while our models
roughly fit the morphology of the ring, they fail to fit the photometry.  
Simple Gaussian ring models are not sufficient to model all features
of the disc, but still can yield some insight into basic disc morphology.



\subsubsection{Insertion of Model Discs}

To enable qualitative forward-modeling of disc structure, we 
inserted simulated discs into a second NICI
dataset with similar target star brightness and field-rotation as the
HD 141569 dataset, but with no detected disc or point sources in the 
field of view.  We selected the NICI Campaign dataset of 
2MASS J04472312-2750358 (henceforth 2M J0447)
taken on 2010-01-10 for this purpose.  2M J0447 has a similar H band
magnitude as HD 141569 (H=7.1 vs. H=6.8). 
The 3122 s dataset was taken with the same
filter set as our primary HD 141569 dataset (1.58 $\mu$m and 1.65
$\mu$m narrowband filters), with the same 60s base exposure time, 
 and had a similar amount of field rotation
as the HD 141569 dataset (58.9$^{\circ}$ for 2M J0447 vs. 58.7$^{\circ}$ for the HD
141569 dataset).   

Model discs at arbitrary flux levels were inserted
into each raw data frame (with single frame disc position angle given by the
parallactic angle at the time of observation).  Data with simulated
discs added were then reduced using the NICI campaign pipeline
described above.   We used the mean ring fit parameters from the
previous section as a baseline model and then altered its parameters
to see if we could better qualitatively reproduce the observed structure.  

\begin{figure}
\includegraphics[width=7in]{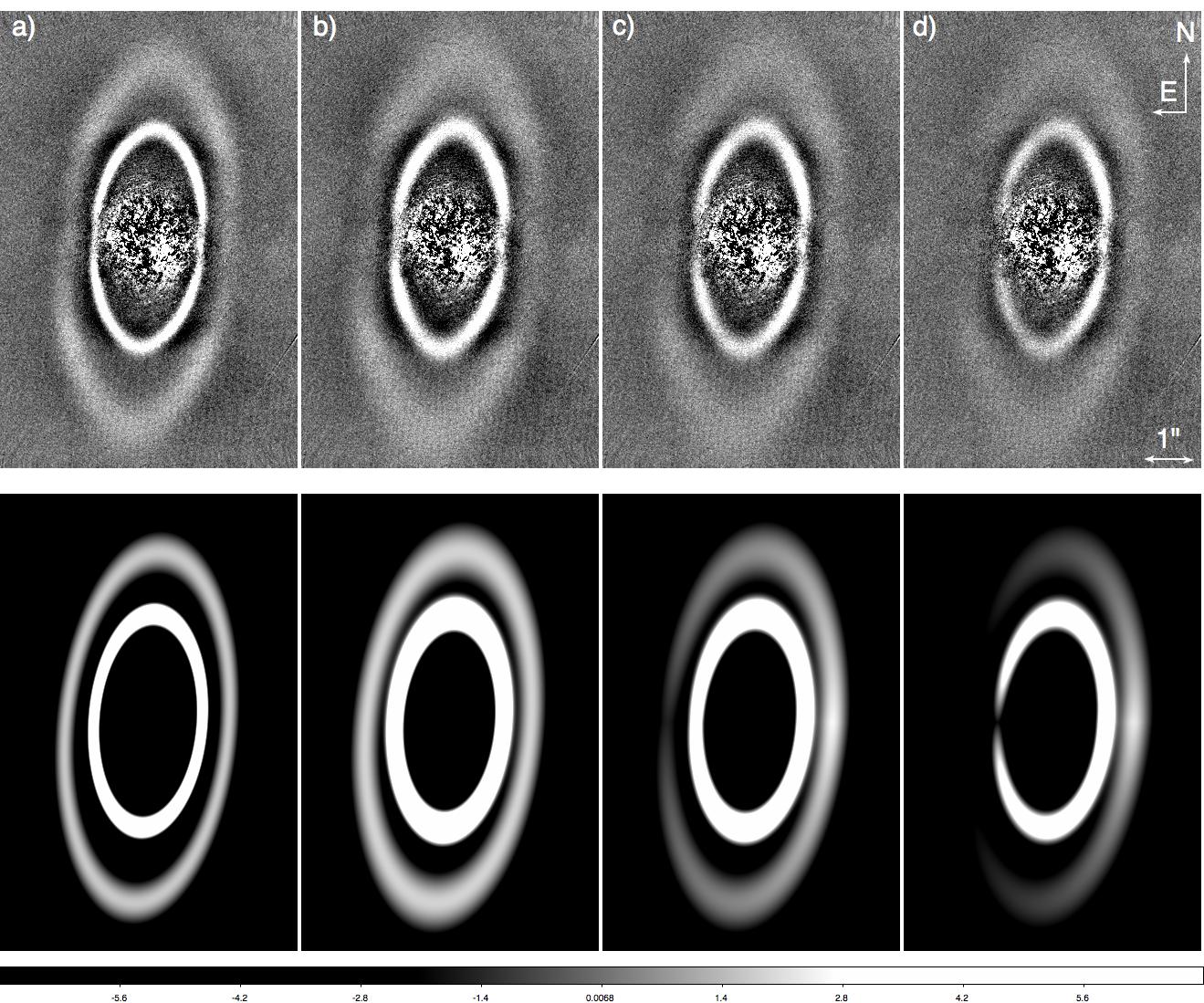}
\caption{\label{fig:selfsubmodeldisk}
Model disc inserted into a NICI campaign dataset (2M0447) with similar 
field rotation as the HD 141569 dataset.  Data with simulated
discs at arbitrary S/N 
added to each data frame were then reduced using the NICI campaign
pipeline.   
For each disc, the flux is arbitrarily scaled from that of the MCMC best fit to the
data in order to ensure a high S/N ratio detection of the disc in this test.
The four cases simulated are, from left to right in the figure: 
a) the MCMC mean ring fit parameters,  b) the MCMC mean ring fit
parameters with the width of both the inner and outer ring increased
by a factor of 1.5 (henceforth model b), 
c) model b with the east side of the image up to 0.5$\times$ fainter
than the west side (henceforth model c) and
d) model b with the east side of the image up to 0.1$\times$ fainter
than the west side (henceforth model d).
Top images show the final reduced image ; bottom images show the
simulated disc inserted into each data frame.}
\end{figure}

We show results for four simulated discs in Fig.~\ref{fig:selfsubmodeldisk}.
We first consider the mean ring fit parameters from the
previous section as a baseline model (henceforth model a).
As per Milli et al. 2012, the ADI reduction process alters the
features of the disc, acting as a filtering algorithm, causing 
narrowing of the disc along the the minor axis and a dark 
self-subtraction region between the two rings.  This is
clearly seen in the far left side panel of
Fig.~\ref{fig:selfsubmodeldisk}, where the retrieved simulated disc is
notably narrower than the original model.  To better reproduce the 
observed disc width, we considered a second model with the  
MCMC mean ring width parameter increased by a factor of 1.5 
(second panel in Fig.~\ref{fig:selfsubmodeldisk}, henceforth model b).   
This model replicates the observed width of the HD 141569 better than
the narrower model and is adopted for all further simulated discs.

Multiple authors note east / west brightness asymmetries in this disc
\citep[][]{Wei99, Mou01, Cla03}, potentially due to the inclination of
the disc from the line of sight.  To test how much such asymmetries would be affected
by an ADI reduction, we also simulated two cases with a simple ring brightness 
asymmetry, where ring brightness varies from the initial model 
as a function of angle $\theta$, with the west side of the disc along
the minor axis as the brightest point and the east side of the disc
along the minor axis as the faintest point.
We considered two cases here: where the east side of the disc at its
faintest is 0.5$\times$ as bright as the west side at its brightest 
(henceforth model c) and where the east side of the disc at its
faintest is 0.1$\times$ as bright as the west side at its
brightest (henceforth model d).  
After ADI processing, we find that the effects of east / west
brightness asymmetries are relatively subtle in the 0.5$\times$ case 
and hard to disentangle from ADI processing artifacts even in this
high S/N case.

\begin{figure}
\includegraphics[width=7in]{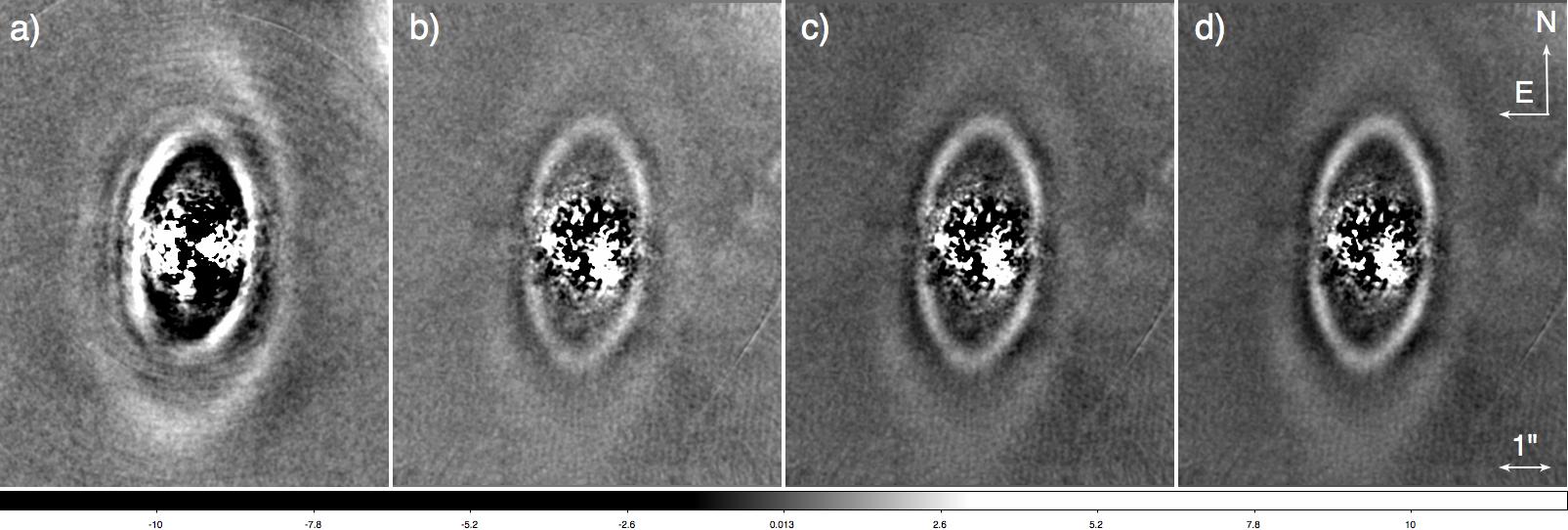}
\caption{\label{fig:selfsubmodeldiskSN}
Model discs inserted into a NICI campaign dataset (2M0447) with similar 
field rotation as the HD 141569 dataset.
From left to right: a) the observed HD 141569 disc for comparison, b)
model disc b (MCMC mean ring fit parameters with the width of both the
inner and outer ring increased by a factor of 1.5) inserted at low S/N
(marginal detection), c) model disc b inserted at S/N approximating
the HD 141569 observation, and d) model disc b inserted at higher S/N.
}
\end{figure}

These simulated discs are clearly detected at higher S/N than the
actual HD 141569 disc detection. Fig.~\ref{fig:selfsubmodeldiskSN}
shows disc model b inserted into the data at S/N ratios more closely
matching the actual detection of the HD 141569 disc (best by-eye S/N match in
panel c), with the HD
141569 detection shown in the left-hand panel for comparison.  
Qualitatively, model b broadly replicates some of the features seen in the HD
141569 -- specifically the observed double ring feature, with
azimuthal asymmetries (potentially due to self-subtraction) around the ring.
However, it does not exactly replicate the observed 
brightness asymmetries in the disc (most notably the northern and southern gaps)
nor does it replicate the arc-like feature seen in between the inner
and outer rings on the east side of the disc.

\begin{figure}
\includegraphics[width=7in]{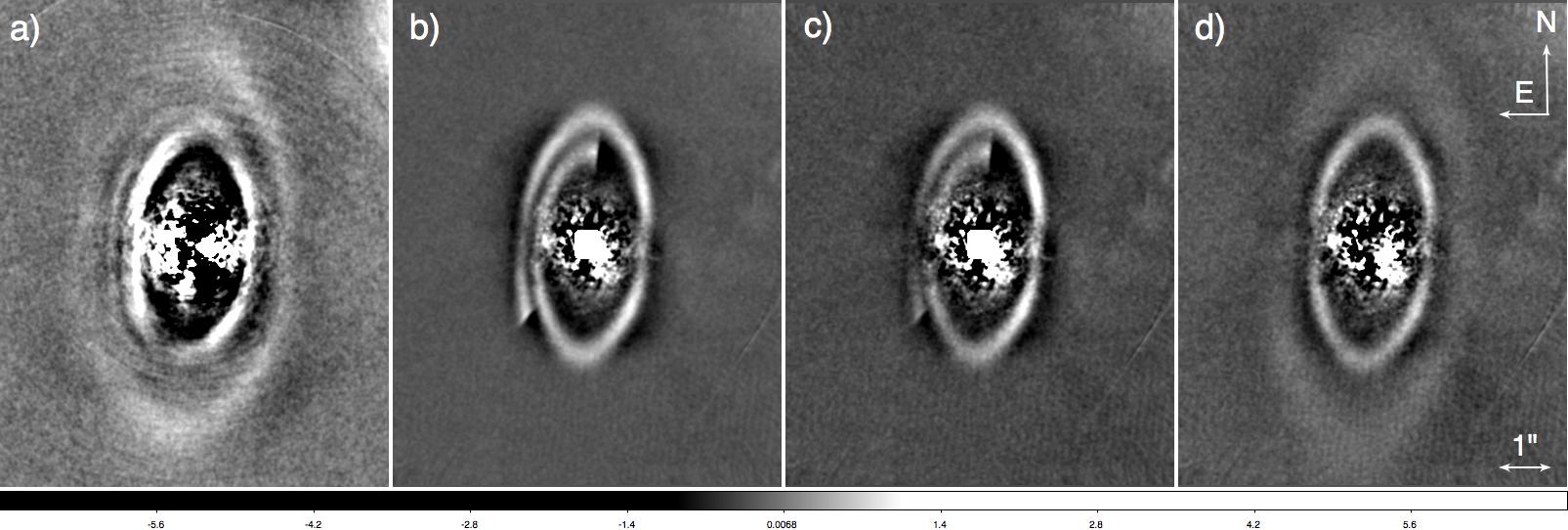}
\caption{\label{fig:selfsubmodelspiral}
Model Archimedean spiral inserted into a NICI campaign dataset (2M0447) with similar 
field rotation as the HD 141569 dataset.
From left to right: a) the observed HD 141569 disc for comparison, 
b) qualitative best match spiral, c) same spiral with a 0.5$\times$
brightness asymmetry imposed, and d) double ring model b shown for comparison.
The width of the spiral feature was set to that of disc model b and
then the Archimedian spiral model was qualitatively adjusted to best
replicate the inner ring / arc structure.  In general, an Archimedean
spiral model does not replicate the observed inner disc structure.}
\end{figure}

Some authors interpret the inner structures in the HD 141569 discs as
potentially due to tightly wound spirals \citep[][]{Cla03}.  This is a
possible explanation for the arc-like feature observed between the
inner and outer ring.  To determine if a spiral might replicate the
observed features of the inner ring better than a gaussian ring model,
we also inserted an Archimedean spiral model into the raw 2M 0447
dataset (more appropriate than a logarithmic spiral, considering the
apparent tight winding). 
The width of the spiral feature was set to that of disc model b and
then the Archimedean spiral model was qualitatively adjusted to best
replicate the inner ring / arc structure.  Results are presented in 
Fig.~\ref{fig:selfsubmodelspiral}, with the observed HD 141569 disc
shown in the left hand panel for comparison.  Panel b shows the best
match spiral model and panel c shows the same model with a 0.5$\times$
brightness asymmetry imposed.  Panel d shows a double gaussian ring
model for comparison.  An Archimedean spiral model with a single peak flux 
does not recreate the observed disc structure.  Although it
also fails to account for all disc features, especially the observed
azimuthal brightness asymmetries as well as the arc-like feature
between the inner and outer spiral / ring features, the double ring model
still qualitatively recreates the inner structure better than the
Archimedean spiral feature.  The reported tightly-wound spiral features are most 
evident in the \citet[]{Cla03} HST optical imaging.
However, as the NICI imaging is at a different wavelength than the
high-resolution optical HST imaging, direct comparison is problematic,
as an accurate comparison would involve significant multi-wavelength
disc modeling.  Thus, we focus here on the morphology found in the
NICI imaging only.

To build an accurate picture of disc structure in this case, we must detangle 
three factors: 1) the S/N of detection of various disc features, 
2) artifacts inevitably introduced by ADI processing, and
3) intrinsic asymmetries due to e.g. disc inclination and actual physical
asymmetries in disc structure.  Since we can do so only broadly 
in the case of HD 141569 (as with an inclination of 51$^{\circ}$, 
the necessary ADI processing inevitably alters the observed disc structure),
we do not interpret disc features in depth.

\subsection{Comparison with HST datasets}

The differences in features recovered between HST and NICI datasets
are likely largely due to differences in sensitivity, resolution,
inner working angle, and data processing methods.  

During the 2011-05-03 NICI observation, we measure a FWHM of
3.8$\pm$0.2 pixels for the 1.58 $\mu$m channel and a FWHM of
4.1$\pm$0.2 pixels for the 1.65 $\mu$m channel.  For the $\sim$18 mas
NICI pixel size, this corresponds to a resolution of 0.07$\arcsec$,
somewhat above the diffraction limit performance for an 8.2 m
telescope at this wavelength (0.05$\arcsec$).  This is a considerable
improvement over the resolution in the HST near-IR imaging of this
disc (diffraction limit of $\sim$0.12$\arcsec$ for a 2.4 m telescope),
but does not reach the 0.05$\arcsec$ resolution achieved in the optical with HST 
\citep[][]{Cla03}.

The NICI imaging is less sensitive to low-surface brightness
features compared to the HST, due to the sky brightness in the near-IR.
Since the HD 141569 disc is so large on the sky ($\sim$4$\arcsec$), most of
the disc lies outside the speckle noise limited portion of the primary
star PSF (corresponding roughly to the inner 0.5$\arcsec$ of the image).  
Thus, sensitivity to faint features does not rely on AO correction but
on the intrinsic background brightness.  This is the likely reason
that we do not recover the faint open spiral features seen in optical
HST imaging by \citet[][]{Cla03}, nor do we recover the lower flux
regions seen between the inner and outer ring/spiral features in HST
imaging.  The deeper ``gap'' we find between the inner and outer
features is a result of disc self-subtraction with ADI processing, 
similar as is found by Milli et al. 2012.

However, NICI can reach much smaller inner working angles than were
reached with the large coronagraphs on HST.  Thus, we have the
clearest picture of the inner 175 AU of the disk, and can trace the 
inner ring / spiral feature over its full 360$^{\circ}$ angular extent 
(although the details of its morphology may have been affected by 
self-subtraction).

\subsection{Limits on Masses of Embedded planets}

As our baseline contrast curve, we adopt the 95$\%$-completeness 
contrast curve calculated for this dataset in the NICI campaign
debris-disc survey paper \citep[]{Wah13a}.  The details of the 
95\%-completeness method are described in the NICI campaign 
pipeline paper \citep[]{Wah13b}.  
The 95\%-completeness technique accounts for self-subtraction losses endemic to ADI and SDI data,
unlike simple measurements based solely on the noise level of the
data.  Briefly, to build the 95\% completeness curve, we first
rereduce the dataset, but with frames mis-rotated to ensure no
positive detections.  This allows us to set a robust false-positive
threshold -- any object ``detected'' in this reduction must be by
definition spurious.  We can also determine a 
nominal 1-$\sigma$ contrast curve from this ``clean reduction''.
The dataset is then rereduced with the correct frame de-rotations 
and with simulated companions 20$\times$ brighter than the 
1-$\sigma$ contrast curve inserted.  The recovered simulated
companions are then inserted into the ``clean reduction'' and scaled 
in intensity until they meet our detection criteria (set by the false-positive threshold).
We adopt the contrast at which 95$\%$ of the simulated companions 
are detected (95$\%$ completeness) as the contrast curve for the star.

\begin{figure}
\begin{tabular}{cc}
\includegraphics[width=3in]{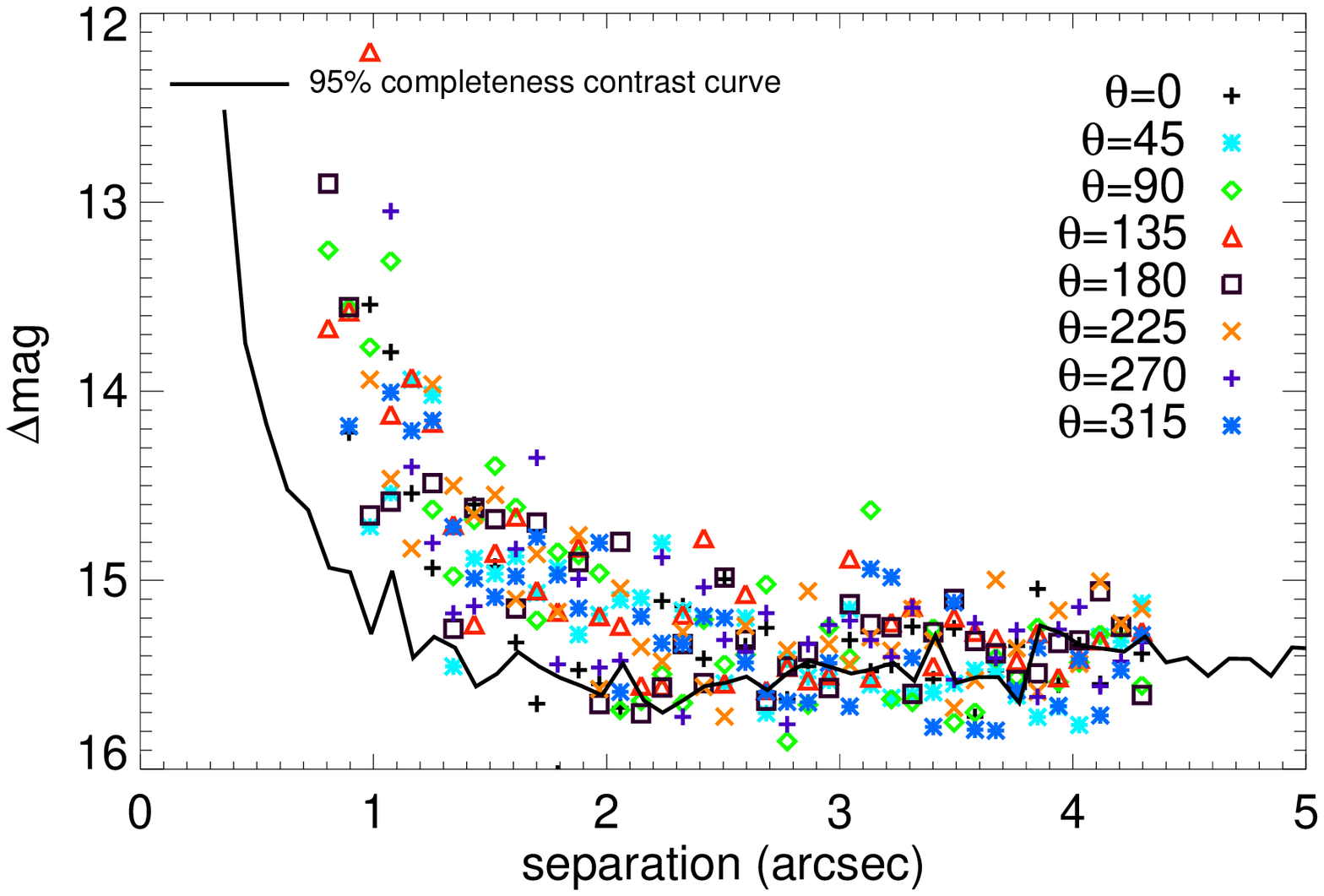} 
& \includegraphics[width=3in]{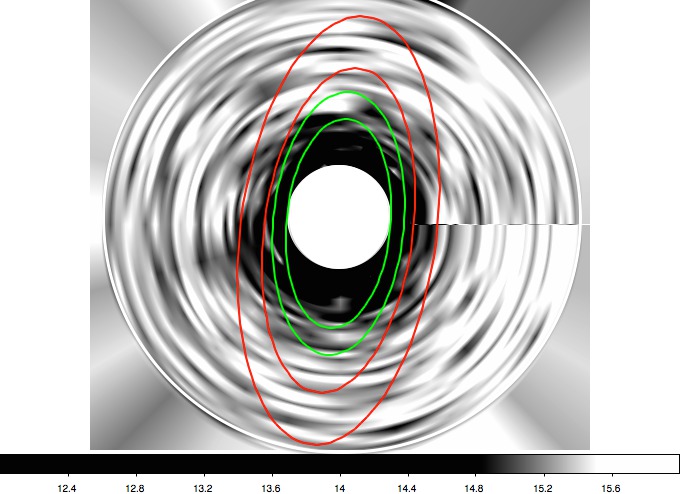} \\
\end{tabular}
 \caption{Left: contrast curves along different angular trajectories 
through the disc.  Right: contrast map obtained by linearly interpolating 
between the contrast curves. Elliptical annuli show the approximate 
position of the inner and outer spiral / ring features.  \label{fig:contrasts}}
\end{figure}

This method works very well for datasets without detected discs or
datasets where the disc has a small angular extent in the field.
However, given the extent and complexity of the HD 141569
disc, self-subtraction is inevitable and flux from the disc remains
even in a mis-rotated reduction.  Thus, to correct for the influence
of the disc, we inserted simulated companions into the raw data 
along different angular trajectories.  Simulated companions were
inserted every 5 pixels from 45 pixels to 240 pixels ($\sim$90-500 AU)
and every 15 degrees in $\theta$.  At each position, simulated companions were
inserted with $\Delta$mag equal to the 95$\%$ contrast curve, and also
from 0.2 to 1 mag brighter than the 95$\%$ contrast curve, with steps of
0.2 mag.   Only one simulated companion was inserted per analysis run, to prevent
interaction / self-subtraction between adjacent objects.  After
insertion and re-reduction, simulated companions were automatically
retrieved using find.pro from the IDL astronomy library.  Significance level of detections
were pinned to the 95$\%$-completeness curve at wide separations, 
where the influence of the disc is negligible.  A contrast map was
produced by linearly interpolating between adjacent curves.  Contrast curves
and the contrast map are presented in Fig.~\ref{fig:contrasts}.

\begin{figure}
\begin{tabular}{cc}
\includegraphics[width=3in]{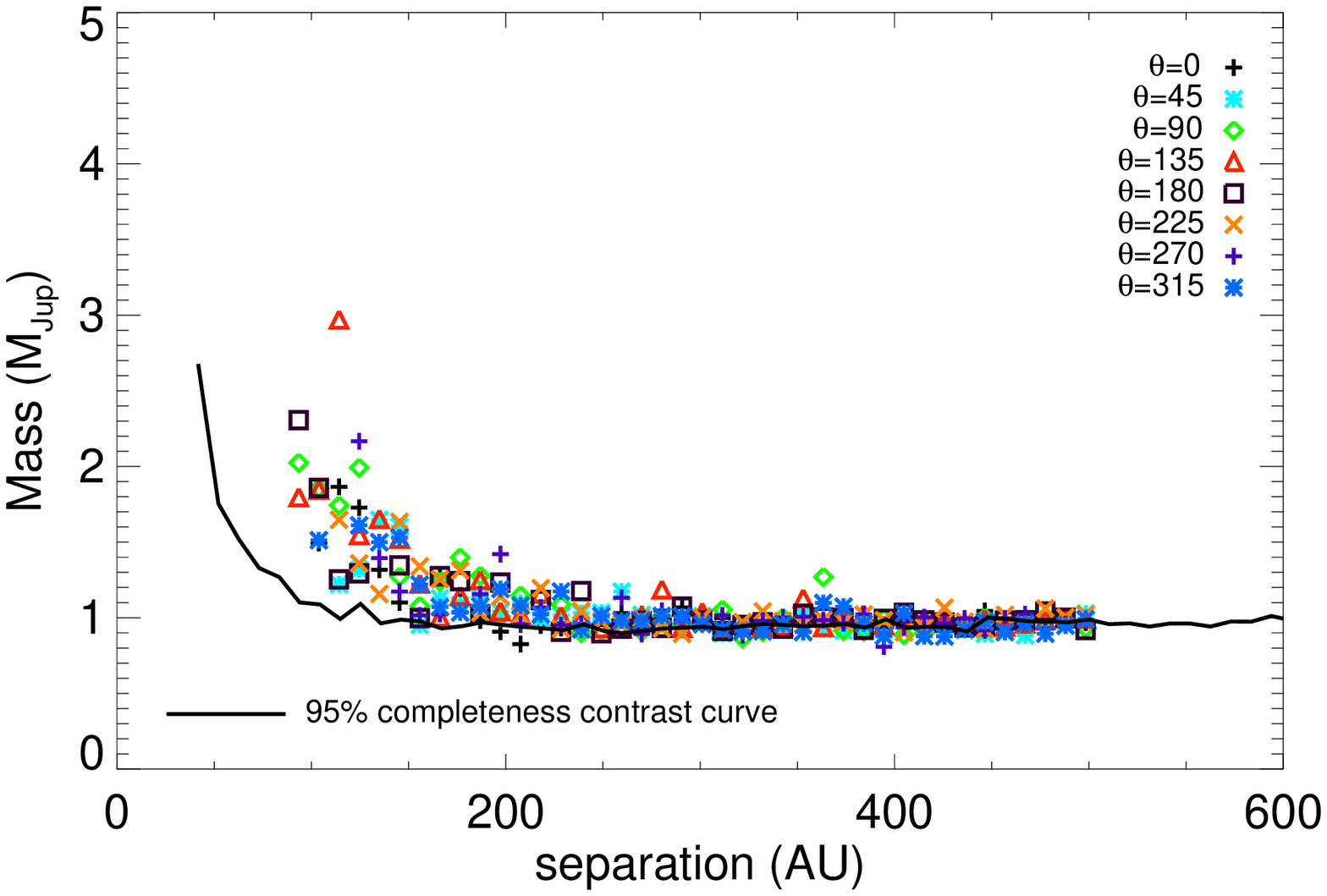} 
& \includegraphics[width=3in]{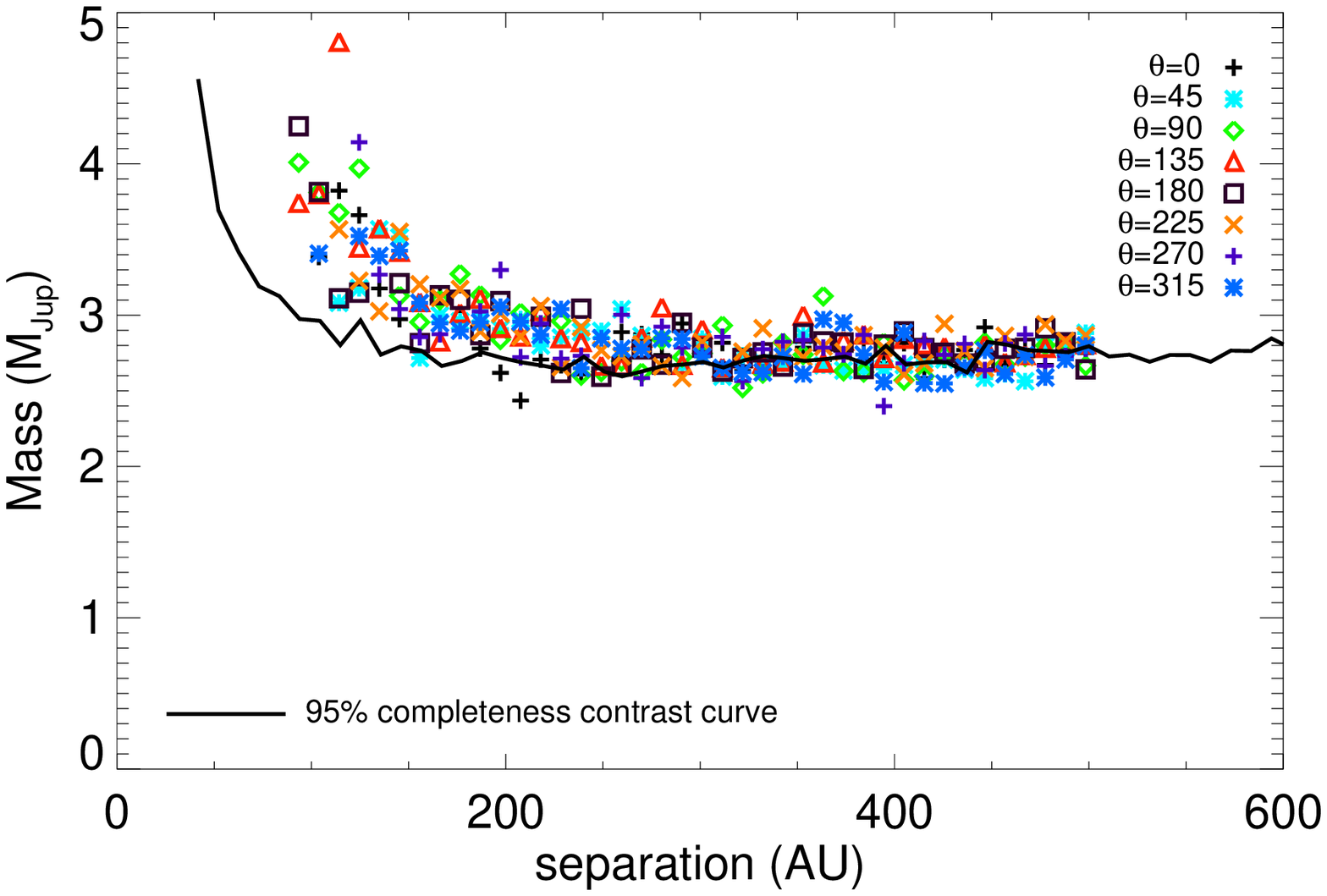} \\
\end{tabular}
 \caption{Minimum detectable mass vs. separation, using the 
Baraffe et al. 2003 COND models (left) and Baraffe et al. 2002
DUSTY models (right), based on the contrast curves from 
Fig.~\ref{fig:contrasts}. \label{fig:masses}}
\end{figure}

We then converted contrast curves to minimum detectable masses using
the hot-start COND and DUSTY models of \citet[]{Bar02, Bar03}, 
adopting an age of 5 Myr and a distance of 116 pc.  
The minimum detectable mass as a function of separation within the disc is
plotted in Fig.~\ref{fig:masses}.  At a very young age of
$\sim$5 Myr, any planets forming or recently formed in the inner or
outer gap of this disc must have formed in situ via gravitational
instability; the system is simply too young for planets forming via
core accretion in the inner disc to have had the time to migrate 
or be scattered to separations $>$100 AU from the host star.  Thus,
hot start evolutionary models are the most appropriate models to
use in estimating our sensitivity to potential planetary companions.
Since it is not clear a companion would be methanated, we consider
both the COND (appropriate for methanated companions) and DUSTY
(appropriate for non-methanated companions) models
and do not consider the SDI reduction for this dataset, as a
non-methanated companion close in would be considerably self-subtracted.
From separations of 200 AU outwards, we can rule out planets with 
mass $\geq$3 M$_{Jup}$.

\section{Discussion}

Numerous modeling efforts have qualitatively reproduced some of the
features in the HD 141569A disc.   Flyby scenarios or encounters with
wide bound companions may be responsible for some of the outer spiral
structure seen in this disc \citep[]{Aug04, Qui05, Ard05, Rec09}, but 
cannot account for the inner gap or spiral features.  The inner
features may be due to secular perturbations from a planet on an
eccentric orbit \citet[]{Wya05}.  Modeling approaches that
combine the effects of a 
fly-by event and one or more embedded planets \citep[]{Ard05, Rec09}, 
have generally been able to reproduce the spiral structures seen with one planet and a fly-by, but
both struggle to reproduce the gap between the inner and outer spiral
/ ring structures.  None of these modeling approaches take into
account the brightness asymmetries found by multiple authors
\citep[]{Wei99, Mou01, Cla03}.

The inner ring / spiral feature is suggestive of the material being
constrained by two planets, one inside its orbit, and one outside,
similar to the shepherd moons observed in Saturns rings
\citep[]{Gol82}.  
To investigate this scenario we ran a suite of N-
body plus test particle simulations with one planet located
at r = 160 AU (where it would be in approximate 2:1 resonance with the
inner edge of the inner ring) and the other at r = 416 AU.
It is then assumed that there is a disc of small test-particles
that extend from r = 130 AU to r = 460 AU. 
We consider 3 cases: both planets
having masses of 2 Jupiter masses and both having no eccentricity, 
both planets having masses of 2 Jupiter masses, with the outer planet 
having an eccentricity of 0.1, 
and then a repeat of the first simulation (no eccentricity) but with a simple drag
force imposed.  We chose a planet mass of 2 Jupiter masses to
simulate, as such a planet might have been missed with the
contrast available with NICI.

\begin{figure}
\begin{tabular}{ccc}
\includegraphics[width=2in]{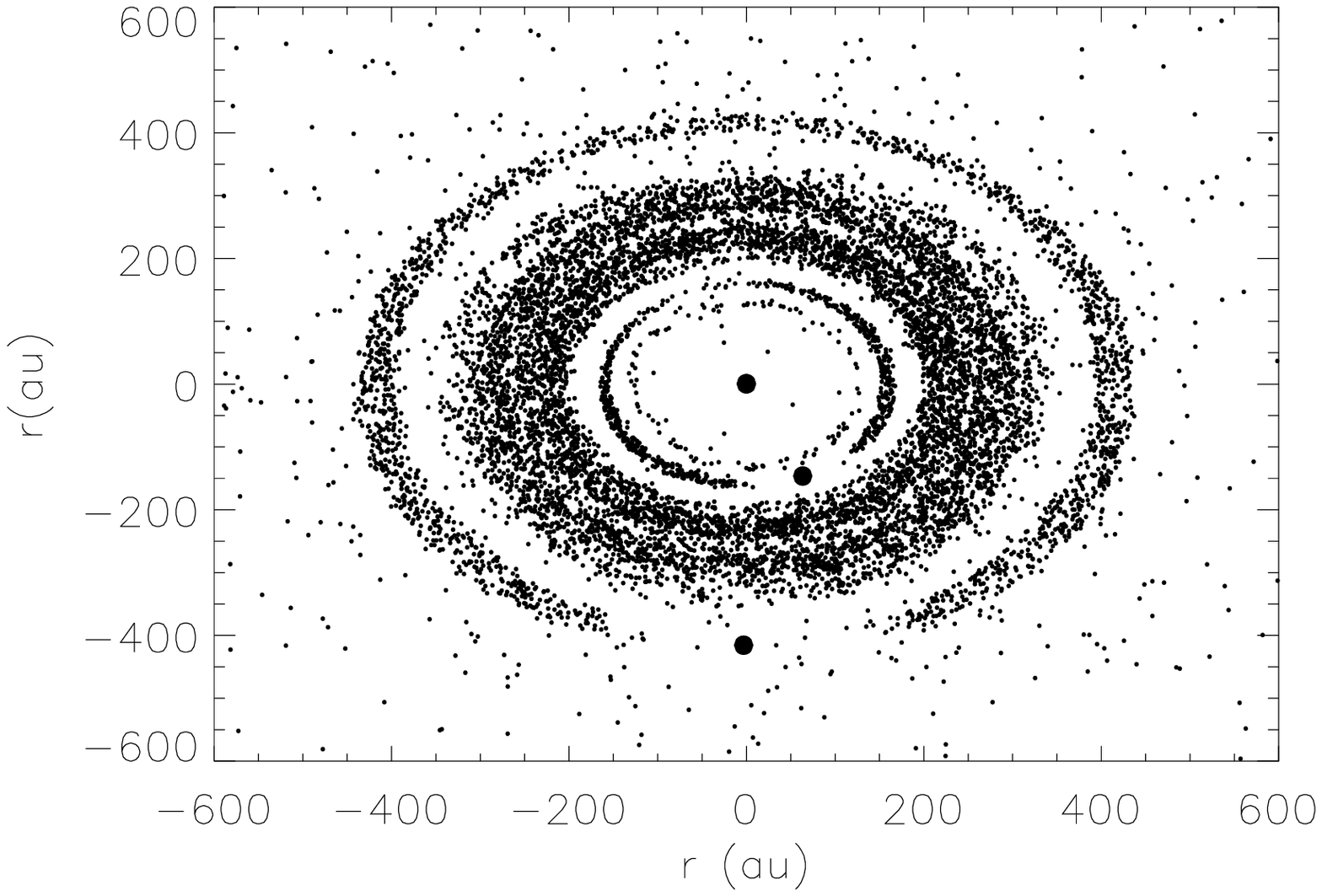} &
\includegraphics[width=2in]{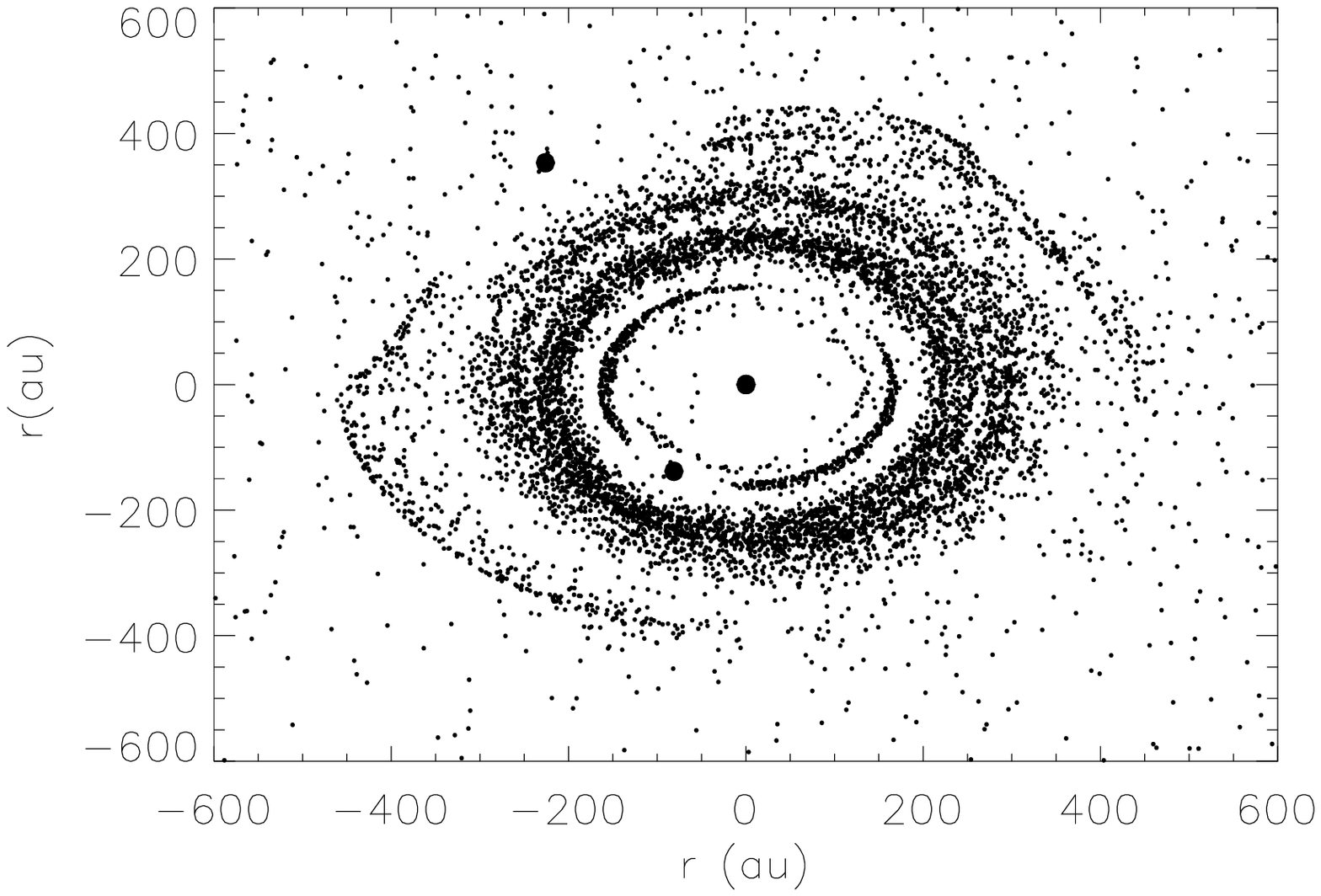} &
\includegraphics[width=2in]{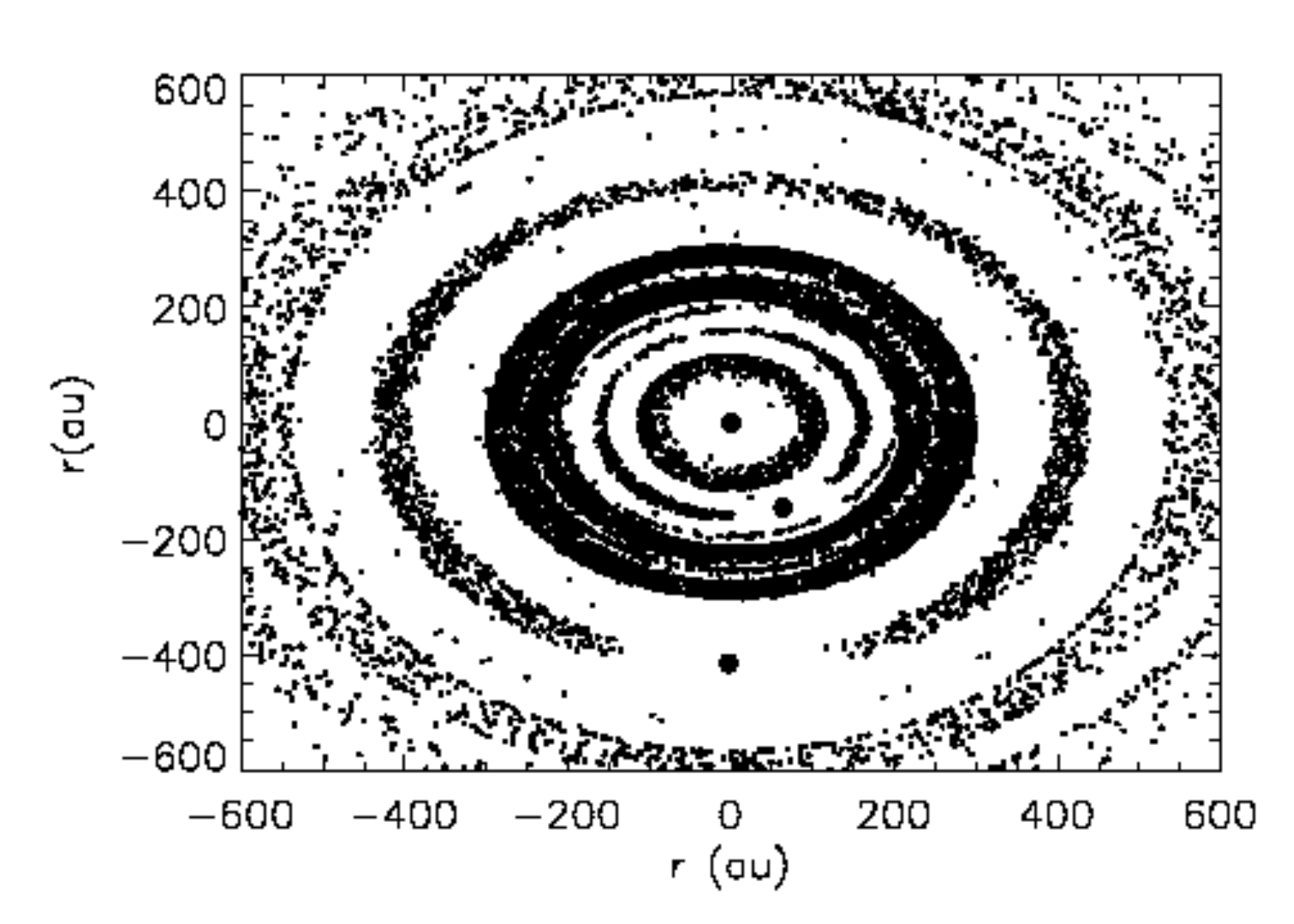} \\
\end{tabular}
\caption{N-body simulation results. 
Left: 2$\times$2 M$_{Jup}$ planets, no
eccentricity, Center: 2$\times$2 M$_{Jup}$ planets, e=0.1
for the outer planet, and Right: 2$\times$2 M$_{Jup}$ planets, no eccentricity, gas drag
included.  The asymmetries seen in the outer ring may be
due to a forming analogue of the Trojan asteroids in our own solar system. \label{fig:models}}
\end{figure}

Each simulation was run for 2.5 Myr, a reasonable timescale for 
planet-formation given the $\sim$5 Myr age of the star.
The results are shown in Figure~\ref{fig:models}.  It's clear that the influence of
these planets can produce a relatively narrow ring with a peak in flux at around $r =
200$ AU and that this ring can have some structure.  When the
eccentricities are zero, however, the structure is
symmetric. Increasing the eccentricity of the outer planet (middle
panel) 
can introduce an asymmetry in the inner ring which is more
consistent with what is observed in HD141569.  Additionally, the outer
ring appears similar to, for example, Jupiter's Trojan asteroids
\citep[]{Mor05} which follow the same orbit, but librate around
the $L_4$ and $L_5$ Lagrangian points.

The right panel illustrates the impact of a weak
drag force due to the presence of some residual gas. A drag force could
influence the collection of particles at the outer planet’s
Lagrange points \citep[]{Cha08} and could act to damp the particles'
eccentricities and inclinations, leading to accelerated grain
growth \citep[]{Pea93}  Although we have not done an extensive study
of the influence of gas drag, the bottom right panel in Figure~\ref{fig:models} 
suggests that a weak drag force does not significantly change the basic
structure, but might act to sharpen the features in the particle disc,
which is  broadly consistent with suggestions in \citet[]{Pea93}.

We should be clear, however, that the goal is not to exactly match
the properties of the rings in HD141569, and we clearly have not
done so.  These simulations are simply intended to be illustrative and 
do suggest that these rings may be indicative of the presence
of two low-mass planets ($<$ 2 M$_{Jup}$), one at about r = 150
AU, and the other at about r = 400 AU. We also show that
including a weak drag force does not significantly influence this
basic conclusion, but could act to sharpen some of the
features in the particle disc. This may be relevant given that the disc is thought
to still have between 67 and 164 M$_{\oplus}$ of gas \citep[]{Thi14}.  

\section{Conclusions}

Our NICI imaging recovers four key 
morphological features in the HD 141569 disc at higher near-IR resolution 
and with a better inner working angle than ever before:  
1) an inner ring / spiral feature. Once deprojected, this feature does not appear circular, 
2) an outer ring / spiral which is considerably brighter on the western side
compared to the eastern side \citet[]{Wei99, Cla03},   
3) an additional arc-like feature between the inner and outer ring
only evident on the east side of the image, which may be part of the
tight spiral structure claimed by \citet[][]{Cla03}.  In the deprojected image,
this feature completes the circle of the west side inner ring, and 
4) an evacuated cavity within 175 AU.  Additionally, we find an offset
of $\sim$4 AU between the inner ring center and the stellar position,
possibly hinting at the presence of unseen companions.
We can recreate some but not all of the structure from N-body
simulations with two embedded planets and argue that the asymmetries 
in the outer ring may trace an extrasolar analgue to the Trojan
asteroids seen in our old solar system.
Indeed, in 3-d modeling of a number of other debris discs, 
Nesvold \& Kuchner (2015) find that 
such features should be common in debris discs
with no planet migration and planets on circular orbits.

\section*{Acknowledgments}

Based on observations obtained at the Gemini Observatory, which is operated by the 
Association of Universities for Research in Astronomy, Inc., under a cooperative agreement 
with the NSF on behalf of the Gemini partnership: the National Science Foundation 
(United States), the National Research Council (Canada), CONICYT (Chile), the Australian 
Research Council (Australia), Minist\'{e}rio da Ci\^{e}ncia, Tecnologia e Inova\c{c}\~{a}o 
(Brazil) and Ministerio de Ciencia, Tecnolog\'{i}a e Innovaci\'{o}n Productiva (Argentina).
B.A.B. was supported by Hubble Fellowship grant HST-HF-01204.01-A
awarded by the Space Telescope Science Institute, which
is operated by AURA for NASA, under contract NAS 5-
26555. This work was supported in part by NSF grants AST-
0713881 and AST- 0709484 awarded to M. Liu, NASA Origins
grant NNX11 AC31G awarded to M. Liu, and NSF grant AAG-1109114 awarded to L. Close.
We thank Alycia Weinberger for providing reduced HST NICMOS images for
comparison with the NICI images and the anonymous referee for useful
suggestions which helped improve the manuscript.

\begin{table*}
 \centering
 \begin{minipage}{140mm}
  \caption{NICI observations of HD 141569 \label{tab:obs}}
  \begin{tabular}{@{}lcccc@{}}
  \hline
  Date & Mode & Number of Images & Total Exp. Time & Total Rotation \\
  &   & & (s) & (degrees) \\
 \hline
2009 Mar 7 & ADI & 20 & 1208 & 10.6 \\
2009 Mar 7 & ASDI & 65 & 3902 & 25.1 \\
2010 Apr 8 & ASDI & 45 & 2701 & 23.6 \\
2011 May 3 & ASDI & 114 & 6844 & 58.7 \\
\hline
\end{tabular}
\end{minipage}
\end{table*}

\begin{table*}
 \centering
 \begin{minipage}{140mm}
  \caption{Ring Fit Parameters\label{tab:params}}
  \begin{tabular}{@{}lcccc@{}}
  \hline
Inner Ring & & & & \\
\hline
  Parameter & Mean & Median & RMS & GR statistic \\
\hline
X offset (pixels) & -1.2 & -1.1 & 0.4 & 1.03 \\
Y offset (pixels) &  1.5 & 1.5 & 1.0 & 1.01 \\
X offset (AU) & -2.5 & -2.3 & 0.8 & 1.03 \\
Y offset (AU) &  3.1 & 3.1 & 2.1 & 1.01 \\
PA (degrees) & -8.9 & -8.9 & 1.3 & 1.01 \\
inclination (degrees) & 44.9 & 44.9 & 0.48 & 1.02 \\
B$_{0}$ & 1.1 & 1.1 & 0.09 & 1.11 \\
r$_{0}$ (pixels) & 117.5 & 117.5 & 1.4 & 1.01 \\
r$_{0}$ (AU) & 245 & 245 & 3.0 & 1.01 \\
$\sigma$ (pixels) & 7.0 & 7.0 & 0.55 & 1.02 \\
$\sigma$ (AU) & 14.6 & 14.6 & 1.1 & 1.02 \\
\hline
Outer Ring & & & & \\
\hline
  Parameter & Mean & Median & RMS & GR statistic \\
\hline
X offset (pixels) & -2.3 & -1.8 & 7.1 & 1.007 \\
Y offset (pixels) &  -5.4 & -5.4 & 8.3 & 1.006 \\
X offset (AU) & -4.8 & -3.8 & 14.8 & 1.007 \\
Y offset (AU) &  -11.3 & -11.3 & 17.3 & 1.006 \\
PA (degrees) & -11.3 & -11.0 & 6.1 & 1.008 \\
inclination (degrees) & 47.3 & 47.0 & 3.3 & 1.016 \\
B$_{0}$ & 0.3 & 0.3 & 0.08 & 1.480 \\
r$_{0}$ (pixels) & 194.5 & 194.6 & 6.4 & 1.008 \\
r$_{0}$ (AU) & 406 & 406 & 13.3 & 1.008 \\
$\sigma$ (pixels) & 21.1 & 19.9 & 5.8 & 1.009 \\
\hline
\end{tabular}
\end{minipage}
\end{table*}

\clearpage

\clearpage

\clearpage

\clearpage

\clearpage

\label{lastpage}

\end{document}